\documentclass[journal]{IEEEtran}

\IEEEpubid{\begin{minipage}{\textwidth}\vspace*{48pt}
     Copyright \copyright~2017
    IEEE. Personal use is permitted.  For any other purposes, permission must
    be obtained from the IEEE by emailing pubs-permissions@ieee.org. This is
    the author's version of an article that has been published in this
    journal. Changes were made to this version by the publisher prior to
    publication.  The final version of record is available at
    \url{http://dx.doi.org/10.1109/TSP.2017.2715001}
  \end{minipage}}

\usepackage[T1]{fontenc}
\usepackage{caption}
\usepackage{subcaption}
\usepackage{placeins}
\usepackage{graphicx}
\usepackage{verbatim}
\usepackage{listings}
\usepackage{url}
\usepackage{array}
\usepackage{hyperref}
\usepackage[nospace,noadjust]{cite}
\usepackage{pgf}
\usepackage[cmex10]{mathtools}
\usepackage{amssymb}
\usepackage{bm}

\usepackage{algorithm}
\usepackage{algpseudocode}

\usepackage{amsthm}

\newtheorem{lemma}{Lemma}

\newtheorem{theorem}{Theorem}

  \def\cC{{\mathcal{C}}}

 \def\cN{{\mathcal{N}}}

   \def\bh{{\mathbf{h}}}
     
 \def\bn{{\mathbf{n}}}   
     
\def\bu{{\mathbf{u}}}  \def\bv{{\mathbf{v}}}

\def\bA{{\mathbf{A}}}  \def\bC{{\mathbf{C}}} 
   \def\bH{{\mathbf{H}}}
\def\bI{{\mathbf{I}}}   
 \def\bN{{\mathbf{N}}}  
   
\def\bU{{\mathbf{U}}} \def\bV{{\mathbf{V}}}  
\def\bY{{\mathbf{Y}}} \def\bZ{{\mathbf{Z}}}

\DeclareMathOperator*{\argmin}{arg\!\min}

\def\complexs{{\mathbb{C}}}

\newcommand{\conj}[1]{\overline{#1}}

\newcommand{\Exp}[1]{\mathrm{E}\left\{#1\right\}}

\newcommand{\T}{\mathsf{T}}
\newcommand{\hsppp}{ {\hspace{0.01in}} }
\newcommand{\rowvec}[1]{\begin{bmatrix}#1\end{bmatrix}^{\T}} % \rowvec{1 & 2 & 3}
\newcommand{\ParState}[1]{\parbox[t]{\dimexpr\linewidth-\algorithmicindent}{%
      \setlength{\hangindent}{\algorithmicindent}
      {#1}\strut}}
\newcommand{\ParStateTwo}[1]{\parbox[t]{\dimexpr\linewidth-\algorithmicindent-\algorithmicindent}{%
      \setlength{\hangindent}{\algorithmicindent}
      {#1}\strut}}
\newcommand{\ParStateZero}[1]{\parbox[t]{\dimexpr\linewidth}{%
      \setlength{\hangindent}{\algorithmicindent}
      {#1}\strut}}

\newcommand{\ignore}[1]{}

\DeclareCaptionLabelSeparator{periodspace}{.\quad}
\graphicspath{{./figures/}}

\title{Noisy Beam Alignment Techniques for Reciprocal MIMO Channels}
\author{Dennis~Ogbe,~\IEEEmembership{Student~Member,~IEEE,} David~J.~Love,~\IEEEmembership{Fellow,~IEEE,} and Vasanthan~Raghavan,~\IEEEmembership{Senior~Member,~IEEE}%
\thanks{D. Ogbe and D.\ J.\ Love are with the School of Electrical and
Computer Engineering, Purdue University, West Lafayette, IN 47906, USA (e-mail: \{dogbe,~djlove\}@purdue.edu).}%
\thanks{Vasanthan Raghavan is with Qualcomm, Inc., Bridgewater, NJ 08807, USA
  (email: vasanthan\_raghavan@ieee.org)}
\thanks{This material is based upon work supported in part by the National Science Foundation under Grant No. CNS-1642982. A version of this paper has been published at the IEEE International Conference
on Acoustics, Speech and Signal Processing (ICASSP) 2017, New Orleans, LA.}
}%

\begin{document}

\maketitle

\begin{abstract}
\noindent
  Future multi-input multi-output (MIMO) wireless communications systems will use beamforming
  as a first-step towards realizing the capacity requirements necessitated by the
  exponential increase in data demands.
  The focus of this work is on beam alignment for time-division duplexing (TDD) systems,
  for which we propose a number of novel algorithms.
  These algorithms seek to obtain good estimates of the optimal
  beamformer/combiner pair (which are the dominant singular vectors of the
  channel matrix). They are motivated by the power method, an iterative
  algorithm to determine eigenvalues and eigenvectors through repeated matrix
  multiplication.  In contrast to the basic power method which considers only
  the most recent iteration and assumes noiseless links, the proposed
  techniques consider information from all the previous iterations of the
  algorithm and combine them in different ways. The first technique ({\em Sequential
  Least-Squares method\/}) sequentially constructs a least-squares estimate of the
  channel matrix, which is then used to calculate the beamformer/combiner pair
  estimate. The second technique ({\em Summed Power method\/}) aims to
  mitigate the effect of noise by using a linear combination of the previously
  tried beams to calculate the next beam, providing improved
  performance in the low-SNR regime (typical for mmWave systems) with minimal
  complexity/feedback overhead. A third technique ({\em Least-Squares Initialized
  Summed Power method\/}) combines the good performance of the first technique at
  the high-SNR regime with the low-complexity advantage of the second technique
  by priming the summed power method with initial estimates from the sequential
  method.
\end{abstract}

 % Simulation results
 %  provide insights on the performance of the proposed algorithms with noisy
 %  links and compare them with similar techniques from the literature. Among
 %  all the schemes studied here, the sequential technique achieves the best
 %  performance as the number of iterations increase or in the high-SNR regime
 %  since the channel matrix estimation step re-orients the beam alignment
 %  problem at every iteration. However, this improvement comes with additional
 %  complexity and feedback overhead.

\begin{IEEEkeywords}
\noindent
  Beam alignment, beamforming, channel reciprocity, TDD, channel estimation,
  massive MIMO, mmWave MIMO, power method.
\end{IEEEkeywords}

\section{Introduction}
\label{sec:intro}

\IEEEPARstart{A}{dvanced} multi-input multi-output (MIMO) systems will be
among the most important technologies to realize the ever-increasing data rate
demands of 5G wireless communication
networks~\cite{andrews_what_2014,boccardi_five_2014}. The two most promising
MIMO applications\footnote{We use the terms massive and mmWave MIMO in the sense
of the common understanding at 3GPP 5G-NR with massive MIMO typically corresponding
to sub-$6$ GHz systems and mmWave MIMO typically corresponding to over-$25$ GHz systems.},
millimeter-wave (mmWave) MIMO~\cite{khan,rappaport,rangan_millimeter-wave_2014,swindlehurst_mmw} and massive
MIMO~\cite{marzetta_noncooperative_2010,ngo_energy_2013,rusek}, rely on utilizing
large beamforming gains to realize the large data rate requirements set for future 5G
networks. In mmWave systems, beamforming will be used to compensate for the
increased path and penetration losses in the 25--100 GHz
band~\cite{5G_whitepaper,3gpp_lte_rel14},
whereas massive MIMO systems will multiplex signals of different users via multi-user
beamforming~\cite{gesbert_mumimo,haardt1} in sub-6 GHz bands.

Many recent works such as~\cite{venkateswaran,torkildson,brady,roh} study the
information theoretic limits of beamforming with practical mmWave hardware
constraints. However, the substantial gains promised by these studies can be
realized only if sufficient channel state information (CSI) is available at the
communication nodes. In current state-of-the-art systems, this information is
acquired by the use of channel sounding sequences and feedback~\cite{love_grassmannian_2003,mukkavilli_beamforming_2003,clerckx3,clerckx4,vasanth_gcom13}.
The use of a large number of antenna elements in mmWave and massive MIMO
systems will make CSI acquisition via the traditional approach
impractical~\cite{medard,hassibi,raghavan_ett}. Further, in mmWave channels with
a relatively small coherence period, it is not possible to {\em simultaneously}
estimate all the elements of the channel matrix due to hardware constraints that
render per-antenna sampling inefficient.

One way to circumvent this problem is to exploit the reciprocal nature\footnote{This
work assumes that the radio-frequency (RF) circuit asymmetries in the uplink and
downlink have been compensated via calibration and hence does not consider these
aspects.} of wireless channels using time-division duplexing (TDD) systems. Channel
reciprocity reduces the
overall resources spent on channel sounding since CSI about the channel in one
direction can be used to adapt to the channel in the reverse direction. Without
readily available channel estimates, communication nodes are forced to obtain
their optimal beamformer/combiner pair by sounding different beams during a
beam alignment phase~\cite{hur_millimeter_2013}. Furthermore, since it is
desirable to minimize the usage of time and power resources of the beam alignment
phase relative to actual data transmission~\cite{medard,hassibi,raghavan_ett}, it
is necessary to employ greedy strategies that maximize the signal-to-noise-ratio
(SNR) during each time slot.

\renewcommand{\arraystretch}{1.2}
\begin{table*}[htbp]
  \centering
  \begin{tabular}{|c|c|c|}
    \hline
    & \textsf{\textbf{Computational Count}} & \textsf{\textbf{Feedback}} \\
    \hline
    BIMA (see~\cite{dahl_blind_2004}) & $k_{\mathsf{max}} \cdot {\cal O}(M)$
                      & - \\ \hline
    BSM (see~\cite{gazor_communications_2010}) & $k_{\mathsf{max}} \cdot {\cal O}(M)$
                      & - \\ \hline
    Sequential Least-Squares & $k_{\mathsf{max}} \cdot {\cal O}(M^3)$
                      &  $k_{\mathsf{max}} \cdot B \cdot (M_{r} + M_{t})$ \\ \hline
    Summed Power & $k_{\mathsf{max}} \cdot {\cal O}(M)$
                      & - \\ \hline
    Least-Squares Initialized Summed Power &
    %\parbox[t]{5cm}{%
           $k_{\mathsf{switch}} \cdot {\cal O}(M^3)$ %\\
           $+ (k_{\mathsf{max}} - k_{\mathsf{switch}}) \cdot {\cal O}(M)$
           %}%
    & $k_{\mathsf{switch}} \cdot B \cdot (M_{r} + M_{t})$ \\ \hline
  \end{tabular}
  \caption{Computational complexity and feedback requirements of
  different beam alignment techniques.}
  \label{"tab:complexity"}
  {\vspace{0.05in}}
  \hrule
\end{table*}

One approach to this goal is to leverage the underlying sparse
structure~\cite{oelayach,schniter,ramasamy,raghavan_ett,raghavan_sparse} or the
directional structure~\cite{alkhateeb,sun,adhikary,raghavan_jstsp,vasanth_gcom16}
of mmWave channels via the use of low-complexity beamforming approaches. The focus
of this work is on another approach that leverages greedy TDD-based beamforming.
Many recent works such as~\cite{dahl_blind_2004,tang_iterative_2005,mandelli_blind_2014,gazor_communications_2010,dahl_intrinsic_2007,raghavan_jstsp} have pursued this approach. The common theme that ties these works is the fact that
repeated conjugation, normalization, and retransmission of an arbitrarily initialized
beamforming vector through a reciprocal MIMO channel (with no noise) is akin to
performing the power
method\footnote{The power method is a result from numerical linear
algebra which provides a simple algorithm to find the dominant eigenvector(s)/eigenspace
of a matrix~\cite{golub_matrix_2012}.} on the channel matrix.

Beam alignment algorithms based on the power method are attractive due to their
simplicity and low computational complexity. However, simple implementations
like the ones proposed
in~\cite{raghavan_jstsp,tang_iterative_2005,dahl_blind_2004} are likely to
perform poorly in the low-SNR
regime~\cite{raghavan_jstsp}. %,gazor_communications_2010}.
Other approaches for finding good beams using the power method have been
proposed in~\cite{gazor_communications_2010}
and~\cite{prasad_joint_2011}. These techniques offer improvements on the
robustness and speed of convergence of the basic power method at the cost of
additional complexity. The main idea behind these improved techniques is to
combine previous estimates of the optimal beams with the received information
during each time slot. In addition to these techniques, recent works such
as~\cite{bengtsson} study the application of the more general Arnoldi iteration
to the beam alignment problem. Furthermore, feedback-based beam alignment
techniques for frequency-division duplexing (FDD) systems, which represent the
majority of currently deployed commercial systems, have been studied
in~\cite{choi_training} and~\cite{noh_feedback}.

Building on~\cite{gazor_communications_2010,prasad_joint_2011,bengtsson}, this
paper presents multiple novel techniques for the TDD MIMO beam alignment
problem in reciprocal channels. These techniques improve upon the performance
of the simple power method-based algorithms, especially in low-SNR environments
which are typical of practical mmWave systems~\cite{raghavan_jstsp}. The first
technique, labeled the {\em sequential least-squares method}, is based on
constructing a least-squares estimate of the channel matrix sequentially using
the previously-used sounding beams. The channel estimates at each iteration can
then be used to compute the next sounding beamformer/combiner pair, which is
exchanged through a feedback\footnote{Due to the small packet overheads, the
  feedback link is assumed to be ideal: error-free and incurring no delay.}
link. The second technique, labeled the {\em summed power method}, does not
require a feedback link
and %improves on the low-SNR performance of the basic power method by computing
computes a normalized running sum of the previous beamformers, thus gaining
greater robustness against noise through averaging.

The first technique achieves better performance in the high-SNR regime at the
cost of additional complexity and feedback\footnote{Nevertheless, the feedback
  link in itself is not onerous given that mmWave links are expected to support
  Gbps rates.} overhead. On the other hand, the second technique achieves
better performance in the low-SNR regime and yet does not need significant
complexity/feedback overhead. However, this technique has deteriorating
performance as the SNR increases due to continued noise averaging. To enjoy the
complementary advantages of both techniques, we propose a third technique,
labeled the {\em least-squares initialized summed power method}, that switches
from the first technique to the second technique after a certain number of
iterations. By appropriately choosing the switching point $k_{\sf switch}$,
significant performance improvement
can be realized in the high-SNR regime with a small increase in feedback and
computational complexity. The motivation behind the third technique is that the
high-SNR performance of a beam alignment algorithm critically depends on the
beam initialization. By choosing this initialization from a scheme that rejects
noise near-optimally, we are able to prime a low-complexity scheme and improve
performance. Thus, the proposed approaches in this paper provide useful
low-complexity solutions for realizing the large beamforming gains of mmWave
systems.

Table~\ref{"tab:complexity"} compares the computational complexity and feedback
requirements of the techniques proposed in this paper with those
from~\cite{dahl_blind_2004} and~\cite{gazor_communications_2010} over a run of
$k_{\sf max}$ iterations of each algorithm. In this table, $M = \max(M_r, M_t)$
with $M_r$ and $M_t$ standing for the receive and transmit antenna dimensions,
respectively and ${\cal O}(\cdot)$ stands for the big-O notation:
$f(x) = {\cal O}(g(x))$ as $x \rightarrow \infty$ if
$\lim_{x \rightarrow \infty} \frac{f(x)}{g(x)} \leq K_u$ for some
$K_u < \infty$. The column labeled ``Computational Count'' lists the
approximate number of complex-valued arithmetic operations required during the
beam alignment phase. The column labeled ``Feedback'' lists the approximate
total number of feedback bits exchanged during the beam alignment phase, where
$B$ is the number of bits per complex-valued element of the beamforming
vectors.

\noindent {\bf \underline{Organization:}} This paper is organized as follows. Section~\ref{sec:system-model} provides
an overview of the system model and sets up the beam alignment techniques
discussed in the rest of the paper. Sections~\ref{sec:seq_npm}-\ref{sec:modif-summ-power}
%and~\ref{sec:summed-power-method}
elaborate on the power methods proposed in this work. Simulation results illustrating
the advantages of the proposed techniques are presented in Section~\ref{sec:sims} with
concluding remarks provided in Section~\ref{sec:conclusion}.

\noindent {\bf \underline{Notations:}} The following notations are used in the
paper. Bold upper-case and lower-case
letters (such as ${\mathbf{A}}$ and ${\mathbf{a}}$) denote matrices and column
vectors, respectively. The operators $\left(\cdot\right)^{\mathsf{T}}$,
$\overline{\left(\cdot\right)}$, $\left(\cdot\right)^*$ and
$\left(\cdot\right)^{\dagger}$ denote matrix transposition, element-wise complex
conjugation, matrix Hermitian transposition and Moore-Penrose pseudoinverse
operations, respectively. $\left\|\cdot\right\|_{2}$ denotes the vector $\ell_2$-norm and
$\left\|\cdot\right\|_{F}$ denotes the Frobenius norm of a matrix.
${\bf x} \sim \mathcal{CN}(\bm{\mu}, \bm{\Sigma})$ denotes a complex Gaussian random vector
with mean $\bm{\mu}$ and covariance matrix $\bm{\Sigma}$. $\mathbb{C}^{n \times m}$,
$\mathbb{C}^n$ and ${\rm E}\{\cdot \}$ stand for the space of $n \times m$ complex
matrices, $n \times 1$ complex vectors and the expectation operator, respectively.

\section{System Model}
\label{sec:system-model}
\begin{figure}[ht]
  \centering
  % 242pt is .96* 252 pt, which is the linewidth
  \resizebox{242pt}{!}{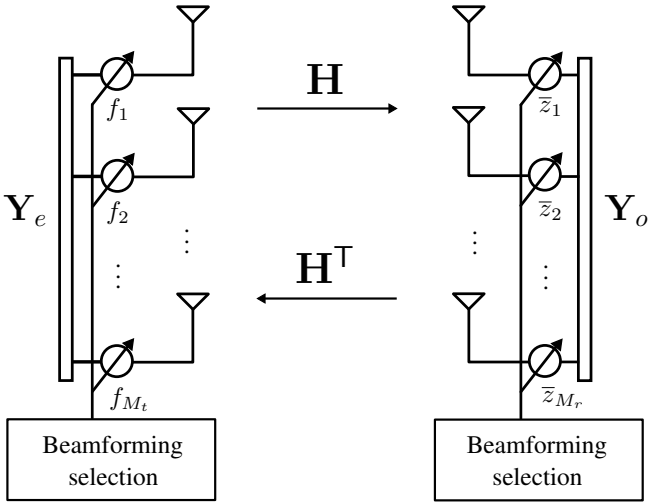}
  \caption{\label{fig:system_model} Communication node 1 transmits data over the
  downlink channel $\mathbf{H}$ to node 2, while node 2 transmits data over the
    uplink channel $\mathbf{H}^{\T}$.
    }
\end{figure}

We consider a multi-antenna communication system such as the one shown in
Fig.~\ref{fig:system_model}, consisting of two transceivers (communication
nodes), with $M_t$ antennas at node 1 and $M_r$ antennas at node 2. The two
nodes communicate over a channel $\mathbf{H}~\in~\mathbb{C}^{M_r \times
  M_t}$. We also assume that $\mathbf{H}$ is reciprocal, i.e.,\ the channel
matrix from node 2 to node 1 (uplink) is the transpose of the channel matrix
from node 1 to node 2 (downlink). For a transmission on the downlink channel,
the transmit data at node 1 is precoded by a unit-norm transmit beamforming
vector
$\mathbf{f}~=~\rowvec{f_1 & f_2 & \hdots & f_{M_t}} \in \complexs^{M_t}$, sent
over the channel, and combined at node 2 with a unit-norm receive combiner
$\mathbf{z}~=~\rowvec{z_1 & z_2 & \hdots & z_{M_r}} \in \complexs^{M_r}$.
Hence, for a data symbol $s_o[k]$ sent on the downlink channel, we obtain the
received symbol
\begin{IEEEeqnarray}{rCl}
  \label{eq:model_received_uplink}
  r_o[k] = \sqrt{\rho_o}\, \mathbf{z}^* \mathbf{H} \mathbf{f} s_o[k] + n_o[k],
\end{IEEEeqnarray}
where $\rho_o$ is the downlink %signal-to-noise ratio (SNR)
SNR and $n_o[k]~\sim~\cC\cN(0, 1)$ is additive Gaussian noise, which we assume
to be independent and identically distributed (i.i.d.) spatially as well as
temporally. Similarly, for a data symbol $s_e[k]$ sent on the uplink channel,
node 1 obtains the received symbol
\begin{IEEEeqnarray}{rCl}
  \label{eq:model_received_downlink}
  r_e[k] = \sqrt{\rho_e}\, \mathbf{f}^{\T} \mathbf{H}^{\T} \conj{\mathbf{z}} s_e[k] + n_e[k].
\end{IEEEeqnarray}

In both~\eqref{eq:model_received_uplink} and~\eqref{eq:model_received_downlink}, we denote
$|\mathbf{z}^* \mathbf{H} \mathbf{f}|^2 = |\mathbf{f}^{\T}
\mathbf{H}^{\T} \conj{\mathbf{z}}|^2$ as the effective channel gain,
which we want to maximize in order to achieve reliable communications and the
highest possible data rates in both directions. We denote the vectors that
achieve this as $\mathbf{f}_{\mathsf{opt}}$ and $\mathbf{z}_{\mathsf{opt}}$, respectively.
It is well-known from~\cite{love_equal_2003,tse_performance_2000} that the
effective channel gain is maximized when $\mathbf{f}$ and $\mathbf{z}$ are the
right- and left-singular vectors of $\mathbf{H}$ corresponding to the largest
singular value of $\mathbf{H}$ and that its maximum achievable value is
$\left\| \mathbf{H} \right\|_{2}^{2} =
\lambda_{\mathsf{max}}\left(\mathbf{H}^{*}\mathbf{H}\right)$.
Further, we assume that neither node has knowledge of the
channel. It is therefore impossible for either node to compute the estimates of
$\mathbf{f}_{\mathsf{opt}}$ and $\mathbf{z}_{\mathsf{opt}}$ using the singular value
decomposition (SVD) of their channel estimate. Instead, as mentioned earlier, these
estimates are obtained iteratively.

In the proposed techniques, both nodes cooperatively determine
$\mathbf{f}_{\mathsf{opt}}$ and $\mathbf{z}_{\mathsf{opt}}$ during a beam
training phase by exploiting the channel's reciprocity property. To model this,
our system operates on a \emph{ping-pong} observation framework, which divides
each discrete channel use into two time slots. During slot 1 (\emph{ping}),
node 1 sends a training symbol to node 2 on the downlink channel
$\mathbf{H}$. During slot 2 (\emph{pong}), node 2 sends a training symbol back
to node 1 on the uplink channel $\mathbf{H}^{\T}$. Since the two nodes are
exchanging training symbols that are known to both sides, we focus on the
received signal vectors after correlating with the known training data. Hence,
the observation at slot 1 (at node 2) during the $k$-th channel use is given as
\begin{IEEEeqnarray}{rCl}
  \label{eq:slot1_obs}
  \mathbf{y}_{o}[k] = \sqrt{\rho_o}\, \mathbf{H}\mathbf{f}[k] + \mathbf{n}_{o}[k].
\end{IEEEeqnarray}
In~(\ref{eq:slot1_obs}), the term $\mathbf{f}[k]$ denotes an estimate of
$\mathbf{f}_{\mathsf{opt}}$ at training phase time-index $k$ and
$\mathbf{n}_{o}[k]~\sim~\cC\cN(\mathbf{0}, \mathbf{I})$ is a complex Gaussian
noise vector of size $M_r$. Due to the reciprocity of the uplink and downlink
channels, the observation at slot 2 (at node 1) is given as
\begin{IEEEeqnarray}{rCl}
  \label{eq:slot2_obs}
  \mathbf{y}_{e}[k] = \sqrt{\rho_e}\, \mathbf{H}^{\T}\conj{\mathbf{z}}[k] + \mathbf{n}_{e}[k].
\end{IEEEeqnarray}
Similar to~\eqref{eq:slot1_obs}, $\rho_e$ denotes the uplink SNR,
$\mathbf{z}[k]$ denotes an estimate of $\mathbf{z}_{\mathsf{opt}}$ at
training phase time-index $k$ and $\mathbf{n}_{e}[k]~\sim~\cC\cN(\mathbf{0}, \mathbf{I})$
is a complex Gaussian noise vector of size $M_t$.

The proposed techniques in this work use all of the ping-pong observations to
determine a good estimate to the optimal beamforming vectors $\mathbf{f}_{\mathsf{opt}}$
and $\mathbf{z}_{\mathsf{opt}}$ in a greedy manner, i.e., each time-index yields
the current choice based on all previously collected observations.

\section{Power Method Using A Sequentially Estimated Channel Matrix}
\label{sec:seq_npm}
\subsection{Batch Least-Squares Estimator}
In the first scheme, since the channel matrix is not known at either node,
the nodes construct a least-squares estimate of $\mathbf{H}$ before each
ping-pong time slot using all of the previous estimates of
$\mathbf{f}_{\mathsf{opt}}$ and $\mathbf{z}_{\mathsf{opt}}$. These estimates
are then used to compute the next state of their beamforming vectors.

In particular, using all observations up to time slot $k$, we can
write~\eqref{eq:slot1_obs} and~\eqref{eq:slot2_obs} in
matrix form as
\begin{IEEEeqnarray}{c}
  \label{eq:slot1_matrix}
  \mathbf{Y}_{o,k} = \sqrt{\rho_{o}}\, \mathbf{H} \mathbf{F}_{k} + \mathbf{N}_{o,k}
\end{IEEEeqnarray}
and
\begin{IEEEeqnarray}{c}
  \label{eq:slot2_matrix}
  \mathbf{Y}_{e,k} = \sqrt{\rho_{e}}\, \mathbf{H}^{\T} \conj{\mathbf{Z}}_k + \mathbf{N}_{e,k}.
\end{IEEEeqnarray}
In~(\ref{eq:slot1_matrix}) and~(\ref{eq:slot2_matrix}),
$\mathbf{F}_{k} = \left[\mathbf{f}[0]\ \mathbf{f}[1]\ \hdots\
  \mathbf{f}[k]\right]$ and
$\mathbf{Z}_{k} = \left[\mathbf{z}[0]\ \mathbf{z}[1]\ \hdots\
  \mathbf{z}[k]\right]$ contain all of the estimates of
$\mathbf{f}_{\mathsf{opt}}$ and $\mathbf{z}_{\mathsf{opt}}$ up to time-index
$k$.  Also,
$\mathbf{Y}_{o,k} = \left[\mathbf{y}_{o}[0]\ \mathbf{y}_{o}[1]\ \hdots\
  \mathbf{y}_{o}[k]\right]$ and
$\mathbf{Y}_{e,k} = \left[\mathbf{y}_{e}[0]\ \mathbf{y}_{e}[1]\ \hdots\
  \mathbf{y}_{e}[k]\right]$ contain all of the observed signal vectors,
respectively. On the other hand,
$\mathbf{N}_{o,k} = \left[\mathbf{n}_{o}[0]\ \mathbf{n}_{o}[1]\ \hdots\
  \mathbf{n}_{o}[k]\right]$ and
$\mathbf{N}_{e,k} = \left[\mathbf{n}_{e}[0]\ \mathbf{n}_{e}[1]\ \hdots\
  \mathbf{n}_{e}[k]\right]$ contain all of the noise vectors, respectively.

Based on this information, node 1 constructs an estimate of the channel
by solving the least-squares problem
\begin{IEEEeqnarray}{c}
  \label{eq:Ho_LS}
  \widehat{\mathbf{H}}_{e,k} =
  \argmin_{\widetilde{\mathbf{H}} {\hspace{0.01in}} \in {\hspace{0.01in}} \complexs^{M_r \times M_t}}
  \left( \left\|\mathbf{Y}_{e,k-1}^{\T} - \sqrt{\rho_{e}}\, \mathbf{Z}_{k-1}^* \widetilde{\mathbf{H}}\right\|^{2}_{F}  \right).
\end{IEEEeqnarray}
Similarly, node 2 constructs an estimate of the channel by solving
\begin{IEEEeqnarray}{c}
  \label{eq:He_LS}
  \widehat{\mathbf{H}}_{o,k} =
  \argmin_{\widetilde{\mathbf{H}} {\hspace{0.01in}} \in {\hspace{0.01in}} \complexs^{M_r \times M_t}}
  \left( \left\|\mathbf{Y}_{o,k} - \sqrt{\rho_{o}}\, \widetilde{\mathbf{H}} \mathbf{F}_{k}\right\|^{2}_{F}  \right).
\end{IEEEeqnarray}
Note that there exists an asymmetry in the time-index between~(\ref{eq:Ho_LS})
and~(\ref{eq:He_LS}). The solutions to these least-squares problems
($\widehat{\mathbf{H}}_{e,k}$ and $\widehat{\mathbf{H}}_{o,k}$) are obtained
using all of the previously observed outputs and beamforming vectors and are
as follows:
\begin{IEEEeqnarray}{c}
  \label{eq:He_batch}
  \widehat{\mathbf{H}}_{e,k} =
  \frac{ \left(\mathbf{Z}_{k-1}^{*} \right)^{\dagger} \mathbf{Y}_{e,k-1}^{\T}}{\sqrt{\rho_{e}}},
  \\
  \label{eq:Ho_batch}
  \widehat{\mathbf{H}}_{o,k} =  \frac{ \mathbf{Y}_{o,k}
  \left( \mathbf{F}_{k}\right) ^{\dagger}}{\sqrt{\rho_{o}}}.
\end{IEEEeqnarray}
In~(\ref{eq:He_batch}) and~(\ref{eq:Ho_batch}), the $(\cdot)^{\dagger}$ operation
stands for the Moore-Penrose pseudoinverse\footnote{Note that the expressions
  in~\eqref{eq:He_batch} and~\eqref{eq:Ho_batch} hold even in the case when
  $k < M_{r}$ since it can be shown that the left pseudoinverse of
  a ``tall matrix'', i.e., a $K_{1} \times K_{2}$ matrix with $K_{1} > K_{2}$
  minimizes $\left\|\bA\bC-\bI\right\|^{2}$, where $\bC$ is optimized over all
  $K_{2} \times K_{1}$ matrices.} of the underlying matrix. Using the definitions
  of the pseudoinverse, we have the following simplifications:
  \begin{align}
  \label{eq:He_batch1}
  \widehat{\mathbf{H}}_{e,k} & = \frac{1}{\sqrt{\rho_{e}}} \cdot \left\{
  \begin{array}{cc}
  {\bf Z}_{k-1} \left( {\bf Z}_{k-1}^{*} {\bf Z}_{k-1} \right)^{-1}
  \mathbf{Y}_{e,k-1}^{\T} & {\sf if} {\hspace{0.05in}} k < M_r  \\
  \left( {\bf Z}_{k-1} {\bf Z}_{k-1}^{*} \right)^{-1} {\bf Z}_{k-1}
  \mathbf{Y}_{e,k-1}^{\T} & {\sf if} {\hspace{0.05in}} k \geq M_r, \\
  \end{array}
  \right. \\
  \label{eq:Ho_batch1}
  \widehat{\mathbf{H}}_{o,k} & = \frac{1}{\sqrt{\rho_{o}}} \cdot \left\{
  \begin{array}{cc}
  \mathbf{Y}_{o,k} \left( {\bf F}_k^* {\bf F}_k \right)^{-1} {\bf F}_k^*
  & {\sf if} {\hspace{0.05in}} k < M_t \\
  \mathbf{Y}_{o,k} {\bf F}_k^* \left( {\bf F}_k {\bf F}_k^* \right)^{-1}
  & {\sf if} {\hspace{0.05in}} k \geq M_t. \\
  \end{array}
  \right.
  \end{align}

Note that the second condition in both~(\ref{eq:He_batch1}) and~(\ref{eq:Ho_batch1})
has been separated (from the first) at the $k = M_r$ and $k = M_t$ cases artificially.
Since the solutions in~(\ref{eq:He_batch}) and~(\ref{eq:Ho_batch}) use all the
underlying data up to time-index $k$, we call this approach the {\em batch
least-squares method}. Once $\widehat{\mathbf{H}}_{e,k}$ and $\widehat{\mathbf{H}}_{o,k}$ have
been estimated, beamforming vector computation follows directly from the
SVD theorem~\cite{golub_matrix_2012,horn_matrix_2012}.
\begin{lemma}
\label{lemma_SVD}
Let the SVD of a matrix $\bA$ be denoted as $\bA = \bU \mathbf{\Sigma} \bV^*$.
We can obtain a multiple of the $i$-th left-singular vector of $\bA$
by multiplying $\bA$ with its $i$-th right-singular vector, i.e., $\bA \bv_i =
\sigma_i \bu_i$. Here, $\sigma_i$ is the $i$-th singular value. Similarly, we
can obtain a multiple of the $i$-th right-singular vector by multiplying $\bA^{*}$
with its $i$-th left-singular vector, i.e., $\mathbf{A}^{*} \bu_i = \sigma_i \bv_i$.
\end{lemma}
Applying Lemma~\ref{lemma_SVD}, we note that node 1 can compute its $k$-th
estimate for $\mathbf{f}_{\mathsf{opt}}$ as
\begin{IEEEeqnarray}{c}
  \label{eq:seq_fk}
  \mathbf{f}[k] = \frac{\widehat{\mathbf{H}}_{e,k}^* \mathbf{z}[k-1]}{\left\|\widehat{\mathbf{H}}_{e,k}^* \mathbf{z}[k-1]\right\|_{2}}.
\end{IEEEeqnarray}
Similarly, applying Lemma~\ref{lemma_SVD}, we note that node 2 obtains its
$k$-th estimate for $\mathbf{z}_{\mathsf{opt}}$ as
\begin{IEEEeqnarray}{c}
  \label{eq:seq_zk}
  \mathbf{z}[k] = \frac{\widehat{\mathbf{H}}_{o,k} \mathbf{f}[k]}{\left\|\widehat{\mathbf{H}}_{o,k} \mathbf{f}[k]\right\|_{2}}.
\end{IEEEeqnarray}

Some comments are in order at this stage.
\begin{enumerate}
\item
We have the following result on error covariance matrices with the batch
estimators.
\begin{theorem}
\label{thm_1}
  If $k \geq \max(M_t, M_r)$, the error covariance matrices of the columns of
  $\widehat{\mathbf{H}}_{e,k}$ and $\widehat{\mathbf{H}}_{o,k}$ under the assumption
  of a channel ${\bf H}$ with independent and identically distributed (i.i.d.) entries
  are given as
  \begin{IEEEeqnarray}{rCl}
    \label{eq:He_batch_cov}
    \mathbf{C}_{e,k} = \frac{1}{\rho_{e}} \left( \mathbf{Z}_{k-1} \mathbf{Z}_{k-1}^* \right)^{-1}
  \end{IEEEeqnarray}
  and
  \begin{IEEEeqnarray}{rCl}
    \label{eq:Ho_batch_cov}
    \mathbf{C}_{o,k} = \frac{1}{\rho_{o}} \left( \mathbf{F}_{k} \mathbf{F}_{k}^* \right)^{-1},
  \end{IEEEeqnarray}
  respectively.
\end{theorem}
For the proof, see Appendix~\ref{appA}.

\item
The proposed algorithm is valid for a general channel matrix ${\bf H}$ and the
i.i.d.\ assumption has been made only in the context of Theorem~\ref{thm_1}.
From Theorem~\ref{thm_1}, we note that the estimation error is monotonically
decreasing in the SNRs, $\rho_e$ and $\rho_o$. This shows that a reasonable
channel estimate can be obtained in the medium- to high-SNR regimes. Nevertheless,
the low-SNR regime is typical in mmWave systems, especially with self-blocking
or blocking due to other humans, vehicles, buildings, foliage, etc.~\cite{raghavan_jstsp,3gpp_lte_rel14}.
Thus, Section~\ref{sec:sims} studies the performance of the different approaches
proposed in this work as a function of the SNR as well as for both i.i.d.\ and
sparse channel models.

\item
While we need $\rho_e$ and $\rho_o$ to compute $\widehat{\mathbf{H}}_{e,k}$ and
$\widehat{\mathbf{H}}_{o,k}$, the beamformer estimates do not depend on these quantities.
Therefore, a mismatched estimate of $\rho_e$ and $\rho_o$ is still sufficient to implement
the proposed scheme.

\item
The computation of $\widehat{\mathbf{H}}_{e,k}$ and ${\bf f}[k]$ at node 1 requires the
feedback of ${\bf z}[k-1]$ from node 2. Similarly, computation of
$\widehat{\mathbf{H}}_{o,k}$ and ${\bf z}[k]$ at node 2 requires the feed forward
of ${\bf f}[k]$ from node 1. While on a first glance this feedback and feed forward
sounds onerous, given the Gbps rates that mmWave systems are expected to realize,
these feedback overheads can be supported on either a lower frequency control/data
channel or on a mmWave control channel. %~\cite{vasanth_patent}.
This feedback/feed forward has to be specified only over a large sub-band
(a component carrier, for example) or on a wideband basis, further reducing the
overhead. Thus, it makes sense to not
dismiss this approach as impractical and study its performance gain relative to
other competing approaches. This is the subject of Section~\ref{sec:sims}. We will
also consider other lower feedback overhead approaches in
Sections~\ref{sec:summed-power-method} and~\ref{sec:modif-summ-power}.

\item
  Throughout this text, we are assuming that the initial transmit beam
  $\mathbf{f}[0]$ is a unit-norm complex random vector.  An alternative
  approach which could be considered for channels with a large line-of-sight
  component would be to initialize $\mathbf{f}[0]$ with an omni-directional
  beam pattern that approximates equal gain in every spatial
  direction. Omni-directional beams have been constructed and used
  in~\cite{hur_millimeter_2013}, but are out of the scope of this work.

\end{enumerate}

The batch least-squares estimators are obtained by computing the Moore-Penrose
pseudoinverse. The complexity in computing these estimators %channel estimates
in~(\ref{eq:He_batch1}) and~(\ref{eq:Ho_batch1}) is limited to the inversion of a
$\widetilde{k} \times \widetilde{k}$ matrix where $\widetilde{k} = \min(M_r, k)$ in
the former case and $\widetilde{k} = \min(M_t, k)$ in the latter case. However,
computation of the matrix to be inverted requires a multiplication count that
scales with $k$ and can hence be onerous.

\subsection{Sequential Least-Squares Estimator (Optimal)}
Following a similar approach to~\cite{kay_fundamentals_1993}, we therefore propose
a sequential algorithm that updates each previous channel estimate based on the
current received signal vector. This approach minimizes computational burden as
well as eliminates the need to store all of the previously received signal %vectors
and beamforming vectors.
Since~\eqref{eq:seq_fk} uses the conjugate transpose of the channel to compute
a new beamformer, we use an algorithm that directly computes an estimate
for $\widehat{\mathbf{H}}_{e,k}^*$ instead of $\widehat{\mathbf{H}}_{e,k}$. This
choice is made here simply to make the derivation of the sequential formulas
more consistent between the two nodes. In this setup, the sequential version
of~\eqref{eq:He_batch} (the channel estimator update) is given as
\begin{IEEEeqnarray}{l}
  \label{eq:He_seq}
  \widehat{\mathbf{H}}_{e,k}^* = \widehat{\mathbf{H}}_{e,k-1}^* + \left(\frac{\conj{\mathbf{y}}_e[k-1]}{\sqrt{\rho_{e}}} - \widehat{\mathbf{H}}_{e,k-1}^* \mathbf{z}[k-1] \right) \mathbf{K}_{e,k}\qquad
\end{IEEEeqnarray}
where
\begin{IEEEeqnarray}{c}
  \label{eq:Ke}
  \mathbf{K}_{e,k} = \frac{\mathbf{z}^*[k-1] \mathbf{C}_{e,k-1}}{1 + \mathbf{z}^*[k-1] \mathbf{C}_{e,k-1} \mathbf{z}[k-1]}
\end{IEEEeqnarray}
and the covariance matrix update is given as
\begin{IEEEeqnarray}{c}
  \label{eq:Sig_e}
  \mathbf{C}_{e,k} = \mathbf{C}_{e,k-1} \left( \mathbf{I} - \mathbf{z}[k-1] \mathbf{K}_{e,k} \right).
\end{IEEEeqnarray}
After obtaining $\widehat{\mathbf{H}}_{e,k}^*$, node 1 uses~\eqref{eq:seq_fk} to obtain
the $k$-th estimate for $\mathbf{f}_{\mathsf{opt}}$. The value of this beamformer then needs to
be fed back to node 2, where it will be used to obtain the next estimate
for $\mathbf{z}_{\mathsf{opt}}$.

At node 2, the same sequential algorithm is used to solve the least-squares
problem, and the update expression for $\widehat{\mathbf{H}}_{o,k}$ becomes
\begin{IEEEeqnarray}{c}
  \label{eq:Ho_seq}
  \widehat{\mathbf{H}}_{o,k} = \widehat{\mathbf{H}}_{o,k-1} + \left( \frac{\mathbf{y}_{o}[k]}{\sqrt{\rho_{o}}} - \widehat{\mathbf{H}}_{o,k-1} \mathbf{f}[k] \right) \mathbf{K}_{o,k}
\end{IEEEeqnarray}
where
\begin{IEEEeqnarray}{c}
  \label{eq:Ko}
  \mathbf{K}_{o,k} = \frac{\mathbf{f}^*[k] \mathbf{C}_{o,k-1}}{1 + \mathbf{f}^*[k] \mathbf{C}_{o,k-1} \mathbf{f}[k]}
\end{IEEEeqnarray}
with the covariance matrix update
\begin{IEEEeqnarray}{c}
  \label{eq:Sig_o}
  \mathbf{C}_{o,k} = \mathbf{C}_{o,k-1} \left( \mathbf{I} - \mathbf{f}[k] \mathbf{K}_{o,k} \right).
\end{IEEEeqnarray}
Node 2 then obtains ${\bf z}[k]$ from~\eqref{eq:seq_zk}, which in turn is fed
back to node 1 to compute ${\bf f}[k+1]$.

We observe that these sequential least-squares (SLS) estimators are only equivalent
to their batch estimators when the beamformer matrices $\mathbf{F}_{k}$ and $\mathbf{Z}_{k}$
are of full column rank. That is, for $k \leq
\text{rank}[\mathbf{H}]$, both
nodes would need to compute their channel estimates using the batch
approach. Theorem~\ref{lem:seq_equivalent} establishes that the sequential
approach is equivalent to using the batch estimator for all $k$.
\begin{theorem}
  \label{lem:seq_equivalent}
  The sequential least-squares estimator $\widehat{\mathbf{H}}_{o,k}^{\mathrm{Seq}}$
  is identical to the batch least-squares estimator
  $\widehat{\mathbf{H}}_{o,k}^{\mathrm{Batch}}$ for $k > r$ if
  $\widehat{\mathbf{H}}_{o,r}^{\mathrm{Seq}} =
  \widehat{\mathbf{H}}_{o,r}^{\mathrm{Batch}}$ where
  $r = \text{rank}[\mathbf{H}]$.
\end{theorem}
\noindent For the proof, see Appendix~\ref{appC}.

Motivated by Theorem~\ref{lem:seq_equivalent}, we propose to initialize
$\mathbf{f}[0]$
as a complex random unit-norm vector. We then use the batch
estimator from~\eqref{eq:He_batch} and~\eqref{eq:Ho_batch} for
$k \leq \text{rank}[\mathbf{H}]$ and switch to the sequential estimator for
$k > \text{rank}[\mathbf{H}]$.  Under these assumptions, the
Gauss-Markov Theorem states that the least-squares estimator is the best linear
unbiased estimator (BLUE) for the channel matrix $\mathbf{H}$~\cite{kay_fundamentals_1993}.
The asymptotic normality property of the least-squares estimator~\cite{eicker_asymptotic_1963}
then shows how our sequential estimates for the
channel matrix converge to its true value. As the channel estimate becomes more
accurate with the number of iterations, steps~\eqref{eq:seq_fk} and
\eqref{eq:seq_zk} essentially perform a two-iteration power method without noise,
which converges at a rate of $\left( \sigma_{1}/\sigma_{2} \right)^{2}$~\cite{golub_matrix_2012}.
The description under Algorithm~\ref{alg:sls-optimal} gives a %step-by-step
succinct summary of this technique (labeled as {\em SLS Estimator (Optimal)})
corresponding to stopping at $k_{\sf max}$ iterations, where $k_{\sf max}$ is
chosen appropriately.

\begin{algorithm}
  \caption{SLS Estimator (Optimal)}
  \label{alg:sls-optimal}
  \begin{algorithmic}
    \State{Initialize $\mathbf{f}[0]$
      as a complex random unit-norm vector.}
    \ForAll{$k = 1, \hdots, k_{\mathsf{max}}$}
    \State\ParState{Node 2 receives $\mathbf{y}_{o}[k-1]$~as in~(\ref{eq:slot1_obs}) and
      gets~${\mathbf{f}}[k-1]$
      from Node 1}
    \If{$k \leq \mathrm{rank}[\mathbf{H}]$}
    \State Node 2 estimates $\widehat{\mathbf{H}}_{o,k-1}$~as in~\eqref{eq:Ho_batch1}
    \ElsIf{$k > \mathrm{rank}[\mathbf{H}]$}
    \State Node 2 estimates $\widehat{\mathbf{H}}_{o,k-1}$~as in~\eqref{eq:Ho_seq}
    \EndIf
    \State Node 2 computes $\mathbf{z}[k-1]$~as in~\eqref{eq:seq_zk}
    {\vspace{0.1in}}
    \State\ParState{Node 1 receives $\mathbf{y}_{e}[k-1]$~as in~(\ref{eq:slot2_obs}) and
      gets~${\mathbf{z}}[k-1]$ from Node 2}
    \If{$k \leq \mathrm{rank}[\mathbf{H}]$}
    \State Node 1 estimates $\widehat{\mathbf{H}}_{e,k}$~as in~\eqref{eq:He_batch1}
    \ElsIf{$k > \mathrm{rank}[\mathbf{H}]$}
    \State Node 1 estimates $\widehat{\mathbf{H}}_{e,k}$~as in~\eqref{eq:He_seq}
    \EndIf
    \State Node 1 computes $\mathbf{f}[k]$~as in~\eqref{eq:seq_fk}
    \EndFor
  \end{algorithmic}
\end{algorithm}

\subsection{Sequential Least-Squares Estimator (Suboptimal)}
For large antenna dimensions as is typical in mmWave systems, it can be
computationally difficult to use the batch estimator for the first $M_{t}$
iterations. In this case, we initialize the sequential least-squares
estimator with an arbitrary initial covariance estimate. With such a choice,
the following result shows that we are guaranteed to asymptotically approach
the batch least-squares estimate.
\begin{theorem}
  \label{lem:seq_non_optimal}
  The sequential least squares estimate
  $\widehat{\mathbf{H}}_{o,k}^{\mathrm{Seq}}$, initialized with
  $\mathbf{C}_{o, 0}~=~\alpha \mathbf{I}$ approaches the batch least-squares
  estimate $\widehat{\mathbf{H}}_{o,k}^{\mathrm{Batch}}$ as
  $\alpha~\rightarrow~\infty$.
\end{theorem}
\noindent For the proof, see Appendix~\ref{app:lem:seq_non_optimal}.

Using Theorem~\ref{lem:seq_non_optimal}, the alternative algorithm (labeled
as {\em SLS Estimator (Suboptimal)}) also requires us to initialize $\mathbf{f}[0]$
as a complex random unit-norm vector. The nodes then transmit this vector
across $\mathbf{H}$ according to~\eqref{eq:slot1_obs} and~\eqref{eq:slot2_obs} and
compute their initial rank-1 channel estimates and beamforming vectors as follows:
\begin{IEEEeqnarray}{rCl}
  \label{eq:seq_asym_init_1}
  \widehat{\mathbf{H}}_{o, 0} & = &
  \frac{\mathbf{y}_{o}[0]\mathbf{f}^{*}[0]}{\sqrt{\rho_{o}}}
 \\
 \label{eq:step2}
 {\bf z}[0] & = &
 \frac{\widehat{\mathbf{H}}_{o,0} \mathbf{f}[0]}
 {\left\|\widehat{\mathbf{H}}_{o,0} \mathbf{f}[0]\right\|_{2}}
 =
 \frac{ {\bf y}_o[0]} { \| {\bf y}_o[0] \|_2 }
 \\
 \label{eq::seq_asym_init_2}
  \widehat{\mathbf{H}}_{e, 1}^{*} & = &
  \frac{\conj{\mathbf{y}}_{e}[0]\mathbf{z}^{*}[0]}{\sqrt{\rho_{e}}} \\
  \label{eq:step3}
   \mathbf{f}[1] & = &
   \frac{\widehat{\mathbf{H}}_{e,1}^* \mathbf{z}[0]}
   {\left\|\widehat{\mathbf{H}}_{e,1}^* \mathbf{z}[0]\right\|_{2}}.
\end{IEEEeqnarray}
The nodes then initialize $\mathbf{C}_{o, 0} = \mathbf{C}_{e, 1} = \alpha {\bf I}$
for an appropriately chosen $\alpha$.
The nodes then use
the sequential formulas~\eqref{eq:He_seq}-\eqref{eq:Sig_o} to estimate their
beamformers. To conclude this section, Algorithm~\ref{alg:sls-suboptimal}
provides a brief summary of
this technique corresponding to $k_{\sf max}$ iterations.

\begin{algorithm}
  \caption{SLS Estimator (Suboptimal)}
  \label{alg:sls-suboptimal}
  \begin{algorithmic}
    \State\ParStateZero{Initialize $\mathbf{f}[0]$
    as a complex random unit-norm vector and obtain
   $\widehat{\mathbf{H}}_{o, 0}$, ${\bf z}[0]$,
   $\widehat{\mathbf{H}}_{e, 1}$ and $\mathbf{f}[1]$ as
   in~(\ref{eq:seq_asym_init_1})-(\ref{eq:step3}).}
    \State Initialize $\mathbf{C}_{o,0} = \mathbf{C}_{e,1} =
    \alpha \mathbf{I}$ for an appropriate $\alpha$.
    \ForAll{$k = 1, \hdots, k_{\mathsf{max}}$}
    \State\ParState{Node 2 receives $\mathbf{y}_{o}[k]$~as in~(\ref{eq:slot1_obs})
      and gets~${\mathbf{f}}[k]$ \linebreak from Node 1}
    \State Node 2 estimates $\widehat{\mathbf{H}}_{o,k}$~as in~\eqref{eq:Ho_seq}
    \State Node 2 computes $\mathbf{z}[k]$~as in~\eqref{eq:seq_zk}
{\vspace{0.1in}}
    \State\ParState{Node 1 receives
      $\mathbf{y}_{e}[k]$~as in~(\ref{eq:slot2_obs}) and gets~${\mathbf{z}}[k]$
      \linebreak from Node 2}
    \State Node 1 estimates $\widehat{\mathbf{H}}_{e,k+1}$~as in~\eqref{eq:He_seq}
    \State Node 1 computes $\mathbf{f}[k+1]$~as in~\eqref{eq:seq_fk}
    \EndFor
  \end{algorithmic}
\end{algorithm}

\section{Summed Power Method}
\label{sec:summed-power-method}
We now propose an alternate approach, labeled the {\em summed power method}, to align
the beams at the two nodes. The main idea behind this scheme is that both nodes
calculate their next beamformers as a function of the running sum of their
previously received vectors, effectively averaging out noise in the estimation
process.
This low-complexity approach adds only one additional vector addition per
iteration at each node when compared to the simple power
method~\cite{dahl_blind_2004,raghavan_jstsp}. Additionally, there is no need
for a feedback link, as neither node needs to have knowledge of the other node's
beamformer.

As described in Section~\ref{sec:system-model}, both nodes exchange training
symbols according to~\eqref{eq:slot1_obs}~and~\eqref{eq:slot2_obs}. However,
instead of simply conjugating and retransmitting their received vector as in
the simple power method, both nodes obtain their next beamformers from a
running sum of all of their previous received vectors. At each time-index $k$,
node 1 computes its next beamformer as
\begin{IEEEeqnarray}{rCl}
  \label{eq:sum:node1-sum}
  \mathbf{f}[k+1] &=&
  \alpha_{k} \left[ \conj{\mathbf{y}}_{e}[k] + \conj{\mathbf{y}}_{e}[k-1] + \cdots + \conj{\mathbf{y}}_{e}[0] \right]
  \\ &=&
  \alpha_{k} {\hspace{0.02in}} \mathbf{s}_{e}[k].
  \label{eq:step4}
\end{IEEEeqnarray}
Similarly, node 2 computes its next beamformer as
\begin{IEEEeqnarray}{rCl}
  \label{eq:sum:node2-sum}
  \mathbf{z}[k+1] &=& \beta_{k} \left[ \mathbf{y}_{o}[k] + \mathbf{y}_{o}[k-1] + \cdots + \mathbf{y}_{o}[0] \right]
  \\
  &=& \beta_{k} {\hspace{0.02in}} \mathbf{s}_{o}[k].
  \label{eq:step5}
\end{IEEEeqnarray}
In~(\ref{eq:step4}) and~(\ref{eq:step5}), $\mathbf{s}_{e}[k]$ and $\mathbf{s}_{o}[k]$
are the state vectors at each node which hold the running sum of the received vectors.
The terms $\alpha_{k}$ and $\beta_{k}$ are normalization factors ensuring the unit-norm
constraint and are given as
\begin{IEEEeqnarray}{rCl}
  \label{eq:sum:alpha_as_norm}
  \alpha_{k} = \frac{1}{\left\| \mathbf{s}_{e}[k] \right\|_{2}}
\end{IEEEeqnarray}
and
\begin{IEEEeqnarray}{rCl}
  \label{eq:sum:beta_as_norm}
  \beta_{k} = \frac{1}{\left\| \mathbf{s}_{o}[k] \right\|_{2}}.
\end{IEEEeqnarray}
Algorithm~\ref{alg:summed} provides an overview of the proposed technique.
\begin{algorithm}
  \caption{Summed Power Method}
  \label{alg:summed}
  \begin{algorithmic}
    \State Initialize $\mathbf{f}[0]$ and $\mathbf{z}[0]$
    as complex random unit-norm vectors.
    \ForAll{$k = 1, \hdots, k_{\mathsf{max}}$}
    \State Node 2 receives $\mathbf{y}_{o}[k-1]$~as in~(\ref{eq:slot1_obs})
    \State Node 2 computes $\mathbf{z}[k]$~as in~(\ref{eq:sum:node2-sum})
    {\vspace{0.1in}}
    \State Node 1 receives $\mathbf{y}_{e}[k-1]$~as in~(\ref{eq:slot2_obs})
    \State Node 1 computes $\mathbf{f}[k]$~as in~(\ref{eq:sum:node1-sum})
    \EndFor
  \end{algorithmic}
\end{algorithm}

For further analysis of the proposed algorithm, it is useful to define the
state-space model of the combined system state: $\mathbf{s}[k]=
\left[ \mathbf{s}_{e}^{\T}[k]\ \mathbf{s}_{o}^{\T}[k] \right]^{\T}$.
A straightforward simplification of $\mathbf{s}[k]$ shows that
\begin{align}
  \label{eq:sum:full-state-vector}
  \mathbf{s}[k] &= \begin{bmatrix} \mathbf{s}_{e}[k] \\
  \mathbf{s}_{o}[k]\end{bmatrix} %\IEEEnonumber
  \\
  &=
  \begin{bmatrix}
    \mathbf{I} & \sqrt{\rho_e} \hsppp \beta_{k-1} \mathbf{H}^{*}\\
    \sqrt{\rho_o} \hsppp \alpha_{k-1} \mathbf{H}  & \mathbf{I}
  \end{bmatrix}
  \mathbf{s}[k-1] + \mathbf{n}[k] \\
  &=
  \prod_{i=0}^{k-1}
  \begin{bmatrix}
    \mathbf{I} & \sqrt{\rho_e} \hsppp \beta_{k-1-i} \mathbf{H}^{*}\\
    \sqrt{\rho_o} \hsppp \alpha_{k-1-i} \mathbf{H} & \mathbf{I}
  \end{bmatrix}
  \mathbf{s}[0]
  \nonumber \\
  & {\hspace{0.1in}}
  + \sum_{\ell=1}^{k} \prod_{j=\ell}^{k-1}
  \begin{bmatrix}
    \mathbf{I} & \sqrt{\rho_e} \hsppp \beta_{k-1+\ell-j} \mathbf{H}^{*} \\
    \sqrt{\rho_o} \hsppp \alpha_{k-1+\ell-j} \mathbf{H} & \mathbf{I}
  \end{bmatrix}
  \mathbf{n}[\ell] %\IEEEnonumber
  \label{eq:citebelow1}
\end{align}
where
\begin{eqnarray}
\mathbf{n}[k] = \left[
\begin{array}{c}
\conj{\mathbf{n}}_{e}[k] \\
\mathbf{n}_{o}[k]
\end{array}
\right].
\end{eqnarray}

Without loss in generality, we can transform an $M_{r} \times M_{t}$ channel
matrix to an $M \times M$ channel matrix by appending zero columns/rows where
$M = \max(M_r, M_t)$. Thus, we restrict attention to square channel matrices.
We can also assume that $\rho_e = \rho_o = \rho$ without loss in generality to
simplify the convergence studies. While establishing a convergence result under
the general Rayleigh fading model appears difficult, we now establish this
under certain restrictions. Nevertheless, numerical studies in
Section~\ref{sec:sims} show that convergence of the summed power method
holds true even for general channel matrix settings. These %restrictions
assumptions (listed as Hypotheses 1-3) are as follows:
\begin{itemize}
\item
{\underline{\bf Hypothesis 1:}} Since convergence studies make more sense
in the high-SNR regime, we assume that $\rho \gg 1$.

\item
{\underline{\bf Hypothesis 2:}} Let ${\bf f}[i] = \left[f_{i,1}, \cdots, f_{i,M}
\right]^{\T}$ and ${\bf z}[i] = \left[z_{i,1}, \cdots, z_{i,M} \right]^{\T}$. We
make the assumptions that as $k$ increases,
$\sum_{i=0}^k f_{i,n} \approx {\sf C}_k$ for all $n$ and
$\sum_{i=0}^k z_{i,m} \approx {\sf C}_k$ for all $m$. In other words, the
statistics of the beamformers remain invariant to the antenna indices at either
node as $k$ increases.

\item
{\underline{\bf Hypothesis 3:}} We consider real-valued, diagonal channel matrices
${\bf H} = {\rm diag}\left( [h_1, \cdots, h_M ] \right)$ with diagonal elements
ordered in non-increasing order. These assumptions can be viewed as restricting all
the signal processing to happen within the bases corresponding to the left- and
right-singular vectors of ${\bf H}$. Also, assume that $h_1 > h_2$ implying a
singular dominant eigen-mode for ${\bf H}$.
\end{itemize}

We now discuss the behavior of the summed power method as $k$ (the number of
iterations) increases under the above assumptions. Under Hypothesis 3, it can
be seen that the optimal beamformers reduce to a scaled version of the first
column of the $M\times M$-dimensional identity matrix, denoted as
$\mathbf{e}_{1}$. Thus, the desired state vector is $\mathbf{s}_{\mathsf{opt}}
= \left[ \alpha \mathbf{e}_{1}^{\T}\ \beta \mathbf{e}_{1}^{\T}
\right]^{\T} = \rowvec{\alpha & 0 & \cdots & 0 & \beta & 0 & \cdots & 0}$ for
some $\alpha$ and $\beta$. The impreciseness in the choice of $\alpha$ and $\beta$
is because the beamforming vector is defined only up to a point on the Grassmann
manifold~\cite{love_grassmannian_2003,mukkavilli_beamforming_2003,vasanth_gcom13}.

Convergence of the summed power method is equivalent
to the limiting behavior/convergence of $\mathbf{s}[k]$
from~(\ref{eq:sum:full-state-vector}) to $\mathbf{s}_{\mathsf{opt}}$.
Lemma~\ref{lem:sum:diagonalization} provides a preliminary result needed to
establish this convergence result.
\begin{lemma}
\label{lem:sum:diagonalization}
Under Hypothesis 3, the state transition matrix %from~(\ref{eq:sum:full-state-vector})
from~(\ref{eq:citebelow1}) is diagonalized by
  \begin{IEEEeqnarray}{rCl}
    \label{eq:sum:diagonalization}
      \mathbf{U}_{k-1} = \begin{bmatrix}
    \sqrt{\frac{\beta_{k-1}}{\alpha_{k-1} + \beta_{k-1} }}\mathbf{I} & \sqrt{\frac{\beta_{k-1}}{\alpha_{k-1} + \beta_{k-1} }}\mathbf{I} \\
    \sqrt{\frac{\alpha_{k-1}}{\alpha_{k-1} + \beta_{k-1} }}\mathbf{I} & -\sqrt{\frac{\alpha_{k-1}}{\alpha_{k-1} + \beta_{k-1} }}\mathbf{I}
  \end{bmatrix}.
\end{IEEEeqnarray}
\end{lemma}
\noindent For the proof, see Appendix~\ref{app:lem:sum:diagonalization}.

Note that ${\bf U}_{k-1}$ is not unitary for general $\alpha_{k-1}$ and
$\beta_{k-1}$. However, we have the following additional result that
simplifies ${\bf U}_{k-1}$.

\begin{lemma}
\label{lem:alphak_betak}
Under Hypotheses 1-3, we can assume that $\alpha_k~\approx~\beta_k$ for large $k$.
Thus, as $k$ increases, $\mathbf{U}_{k-1}$ converges to
\begin{IEEEeqnarray}{rCl}
  \label{eq:sum:U}
  \widetilde{\mathbf{U}} = \frac{1}{\sqrt{2}}
  \begin{bmatrix}
    \mathbf{I} & \mathbf{I} \\
    \mathbf{I} & -\mathbf{I}
  \end{bmatrix}.
\end{IEEEeqnarray}
\end{lemma}
\noindent For the proof, see Appendix~\ref{app:lem:alphak_betak}.

We now have the following main result.
\begin{theorem}
  \label{lem:sum:sing_val_conv}
  Under Hypotheses 1-3, we have that ${\bf s}[k] \rightarrow {\bf s}_{\sf opt}$
  as $k$ increases.
\end{theorem}
\noindent For the proof, see Appendix~\ref{app:lem:sum:sing_val_conv}.

The results of Section~\ref{sec:sims} will show that these results hold for
more general channel models and are not restricted to satisfaction of
Hypotheses 1-3.
In addition, Section~\ref{sec:modif-summ-power} presents two modifications to
the summed power method which aim to improve performance over a wider range
of SNRs while maintaining low computational complexity.

\section{Least-Squares Initialized Summed Power Method}
\label{sec:modif-summ-power}
We now consider a refinement that trades off the advantages of both the
approaches in Sections~\ref{sec:seq_npm} and~\ref{sec:summed-power-method}
in terms of complexity, feedback and performance.
The main motivation behind this approach is the observation that the
performance of a beam alignment algorithm critically depends on how
${\bf f}[0]$ (or ${\bf z}[0]$) is initialized. When ${\bf f}[0]$ is
initialized as a complex random unit-norm vector, we rely on multiple
iterations over the channel to re-align this choice towards the singular
vectors of the channel. Depending on the approach used for alignment as
well as the SNR on the downlink and uplink, the beam alignment algorithm
could take a substantial number of iterations to improve the effective
channel gain.

In this context, we note that the (sequential/batch) least-squares approach
from Section~\ref{sec:seq_npm} achieves good performance in the high-SNR regime
by optimally estimating the channel matrix over every iteration and re-aligning
the alignment problem at every step. However, this gain comes at the cost of
complexity and feedback overhead of the algorithm. On the other hand, at low-SNR,
averaging over the noise results in significant performance improvement with the
summed power method from Section~\ref{sec:summed-power-method}, which is a
low-complexity/feedback overhead scheme.

These observations suggest that the two approaches can be married together,
which is the focus of the {\em least-squares initialized summed power (LISP)
  method}.  In this method, both nodes ``prime'' their beamformers using either
the batch/sequential least-squares method for the first $k_{\sf switch}$
iterations,
after which they switch to the summed power method. In particular, we have the
following description in Algorithm~\ref{alg:sint} for the proposed technique with
the sequential least-squares initialization. The switching point $k_{\sf switch}$
can be chosen in multiple ways. Specific choices for $k_{\sf switch}$ include
$\min(M_r, M_t)$, $\max(M_r, M_t)$ or via some optimality studies as in Sec.~\ref{sec:sims}.

\begin{algorithm}
  \caption{Least-squares Initialized Summed Power Method}
  \label{alg:sint}
  \begin{algorithmic}
    \State\ParStateZero{Initialize $\mathbf{f}[0]$ %and $\mathbf{z}[-1]$
    as a complex random unit-norm vector and obtain
   $\widehat{\mathbf{H}}_{o, 0}$, ${\bf z}[0]$,
   $\widehat{\mathbf{H}}_{e, 1}$ and $\mathbf{f}[1]$ as
   in~(\ref{eq:seq_asym_init_1})-(\ref{eq:step3}).}
    \State Initialize $\mathbf{C}_{o,0} = \mathbf{C}_{e,1} =
    \alpha \mathbf{I}$ for an appropriate $\alpha$.
    \ForAll{$k = 1, \hdots, k_{\mathsf{max}}$}

    \If{$k \leq k_{\sf switch}$}
    \State\ParStateTwo{Node 2 receives $\mathbf{y}_{o}[k]$~as in~(\ref{eq:slot1_obs})
    and gets~${\mathbf{f}}[k]$\linebreak from Node 1}
    \State Node 2 estimates $\widehat{\mathbf{H}}_{o,k}$~as in~\eqref{eq:Ho_seq}
    \State Node 2 computes $\mathbf{z}[k]$~as in~\eqref{eq:seq_zk}
    %\ElsIf{$k = k_{\sf switch}$}
    \ElsIf{$k > k_{\sf switch}$}
    \State Node 2 receives $\mathbf{y}_{o}[k-1]$~as in~(\ref{eq:slot1_obs})
    \State Node 2 computes $\mathbf{z}[k]$~as in~(\ref{eq:sum:node2-sum})
    \EndIf
 {\vspace{0.1in}}
    \If{$k \leq k_{\sf switch}-1$}
    \State\ParStateTwo{Node 1 receives $\mathbf{y}_{e}[k]$~as in~(\ref{eq:slot2_obs})
    and gets~${\mathbf{z}}[k]$ \linebreak from Node 2}
    \State Node 1 estimates $\widehat{\mathbf{H}}_{e,k+1}$~as in~\eqref{eq:He_seq}
    \State Node 1 computes $\mathbf{f}[k+1]$~as in~\eqref{eq:seq_fk}
    \ElsIf{$k > k_{\sf switch}-1$}
    \State Node 1 receives $\mathbf{y}_{e}[k]$~as in~(\ref{eq:slot2_obs})
    \State Node 1 computes $\mathbf{f}[k+1]$~as in~(\ref{eq:sum:node1-sum})
    \EndIf

    \EndFor
  \end{algorithmic}
\end{algorithm}

\section{Numerical Studies}
\label{sec:sims}
In this section, we present performance comparisons of the proposed schemes
obtained via Monte Carlo experiments. We first present results on the convergence
properties of the different techniques under varying conditions. We then present
the impact of an increase in $M_t$ on the performance of these schemes.

\subsection{Convergence Studies}
We study two variants of the proposed sequential least-squares technique from
Section~\ref{sec:seq_npm}: ``SLS (Optimal)'' and ``SLS (Suboptimal).''  The
first variant %, denoted as ``SLS - optimal'' in the plots,
computes the batch least-squares estimator for the first $M_{r}$ (or $M_t$)
iterations before switching to the sequential version after that. The second
variant %, denoted as ``SLS - suboptimal'' in the plots,
relies on the result from Theorem~\ref{lem:seq_non_optimal} to be
computationally efficient and to avoid having to compute the batch estimator.
It is initialized with $\alpha = 1000$ and uses the sequential estimator
starting at the first iteration. We also study the performance of the iterative
solutions based on the summed power method from
Section~\ref{sec:summed-power-method} and the least-squares initialized summed
power method with $k_{\sf switch} = \max(M_r, M_t)$ from
Section~\ref{sec:modif-summ-power}. These approaches are denoted as ``Summed
Power'' and ``LISP'' in the plots, respectively.

In terms of performance benchmarking, we consider the one-dimensional versions
of the techniques proposed in~\cite{dahl_blind_2004} and~\cite{gazor_communications_2010}.
The algorithm from~\cite{dahl_blind_2004} is called {\em Blind Iterative MIMO
Algorithm (BIMA)} by the authors and is denoted as ``BIMA'' in the plots here.
The algorithm from~\cite{gazor_communications_2010} is called {\em Best Singular
Mode (BSM) estimation} by the authors and is denoted as %``Gazor/AlSuhaili''
``BSM'' in the plots here. The value of the design parameter $\mu$ for the BSM
algorithm from~\cite{gazor_communications_2010} is set to $1.5k$ where $k$ is
the time-index.

\begin{figure*}
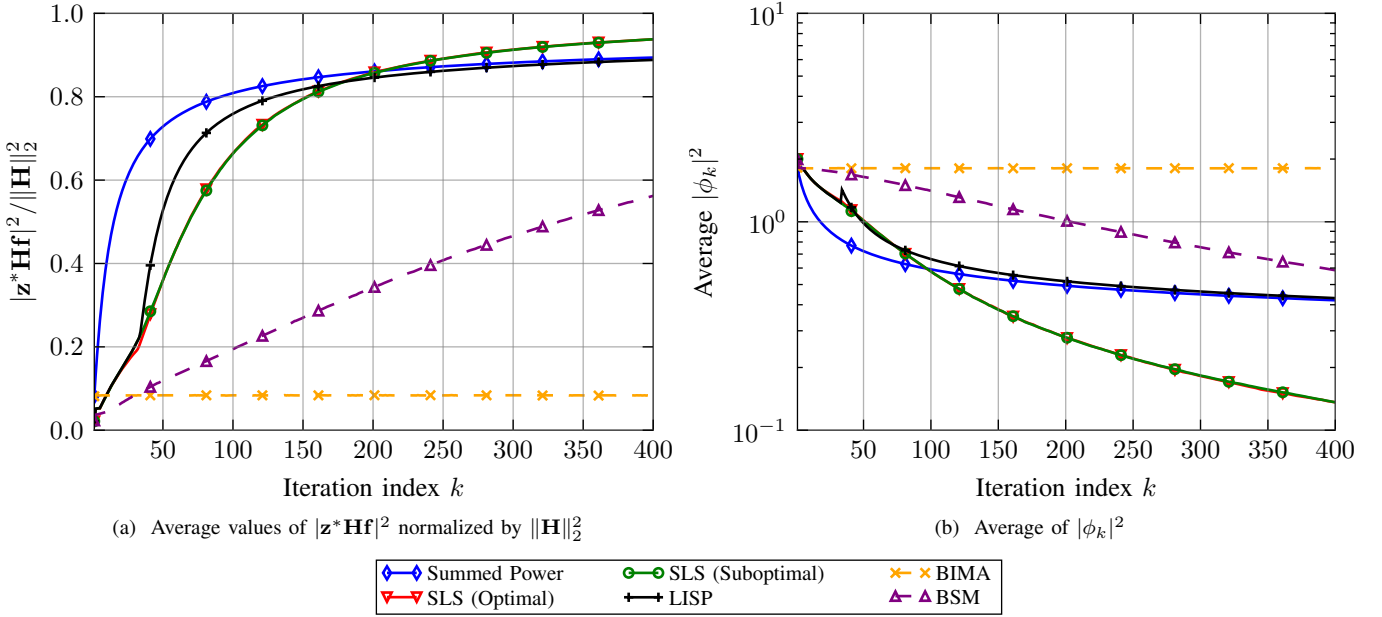

  \centering
  \begin{subfigure}[t]{.5\textwidth}
    \input{./plots/gain_m10dB.pgf}
    \caption{\label{fig:gains_m10dB} Average values of
      $|\mathbf{z}^* \mathbf{H} \mathbf{f}|^2$ normalized by
      $\|\mathbf{H}\|_2^2$}
  \end{subfigure}\hfill
  \begin{subfigure}[t]{.5\textwidth}
    \input{./plots/angle_m10dB.pgf}
    \caption{\label{fig:angle_m10dB} Average of $|\phi_k|^2$}
  \end{subfigure}
      \par\medskip
  \begin{subfigure}[b]{\textwidth}
    \centering\input{./plots/legend_3col.pgf}
  \end{subfigure}
  \caption{\label{fig:m10dB} Results for the i.i.d.\ channel model at
  $\rho_e = \rho_o = -10$ dB with $M_r = 4, M_t = 32$
  }
\end{figure*}

In Figs.~\ref{fig:m10dB}-\ref{fig:20dB}, we compare
the performance of these six schemes at different SNR values with $M_r = 4$
and $M_t = 32$ (corresponding to a downlink channel matrix ${\bf H}$ of
dimensions $4 \times 32$). The channel matrix ${\bf H}$ has i.i.d.\ entries. In
particular, Figs.~\ref{fig:gains_m10dB} and~\ref{fig:angle_m10dB} show the
results for uplink and downlink SNR values of $-10$ dB, whereas,
Figs.~\ref{fig:gains_0dB} and~\ref{fig:angle_0dB}, and
Figs.~\ref{fig:gains_20dB} and~\ref{fig:angle_20dB} provide similar plots for
an SNR of $0$ dB and $20$ dB, respectively. These SNR values are expected to be
typical of low-, medium- and high-SNR regimes, respectively.

Practical mmWave channels are expected to be sparser~\cite{raghavan_sparse,raghavan_jstsp}
than i.i.d.\ channels. In this context, Fig.~\ref{fig:m10dB_sparse} illustrates the
performance of the same set of six schemes in a sparse MIMO channel model with
$\lambda/2$ spaced uniform linear arrays (ULAs) at both ends corresponding to
$M_r = 4$ and $M_r = 32$. Both downlink and uplink SNRs are assumed to be $-10$ dB
and $f_{\sf c} = 28$ GHz is used. The channel is made of $K = 3$ dominant clusters
with one path per cluster (hence the channel matrix ${\bf H}$ is rank-deficient). The
angles of arrival and departure are assumed to be in the azimuth plane and uniformly
distributed in a $120^{\sf o}$ angular spread at both ends. Rayleigh fading is assumed
for the path gains. Such a model is commonly used in mmWave system analysis (see~\cite{raghavan_jstsp}
and references therein for details).

We study two metrics capturing the performance of these six schemes: i) the
instantaneous effective channel gain $|\mathbf{z}^*[k] \mathbf{H} \mathbf{f}[k]|^2$
at time-index $k$, and ii) the angle between the true singular vector
$\mathbf{f}_{\mathsf{opt}}$ and its estimate $\mathbf{f}[k]$, given as
\begin{IEEEeqnarray}{rCl}
  \label{eq:sim_angle}
  \phi_k = \text{cos}^{-1}\left(|\mathbf{f}_{\mathsf{opt}}^* \mathbf{f}[k]|\right),
\end{IEEEeqnarray}
and measured in radians. Note that $\phi_k$ equivalently captures the chordal
distance between $\mathbf{f}_{\mathsf{opt}}$ and $\mathbf{f}[k]$. In order to
average results over different channel realizations, we normalize the
effective channel gain by $\|{\mathbf H}\|_2^2 = \lambda_{\sf max}({\bf H}^* {\bf H})$.
Fast convergence of the algorithm is then equivalent to fast convergence of
the normalized instantaneous effective channel gain to 1.

\begin{figure*}
  \centering
  \begin{subfigure}[t]{.5\textwidth}
    \input{./plots/gain_0dB.pgf}
    \caption{\label{fig:gains_0dB} Average values of
      $|\mathbf{z}^* \mathbf{H} \mathbf{f}|^2$ normalized by
      $\|\mathbf{H}\|_2^2$}
  \end{subfigure}\hfill
  \begin{subfigure}[t]{.5\textwidth}
    \input{./plots/angle_0dB.pgf}
    \caption{\label{fig:angle_0dB} Average of $|\phi_k|^2$}
  \end{subfigure}
      \par\medskip
  \begin{subfigure}[b]{\textwidth}
    \centering\input{./plots/legend_3col.pgf}
  \end{subfigure}
  \caption{\label{fig:0dB} Results for the i.i.d.\ channel model
    at $\rho_e = \rho_o = 0$ dB with $M_r = 4, M_t = 32$}
\end{figure*}

\begin{figure*}
  \centering
  \begin{subfigure}[t]{.5\textwidth}
    \input{./plots/gain_20dB.pgf}
    \caption{\label{fig:gains_20dB} Average values of
      $|\mathbf{z}^* \mathbf{H} \mathbf{f}|^2$ normalized by
      $\|\mathbf{H}\|_2^2$}
  \end{subfigure}\hfill
  \begin{subfigure}[t]{.5\textwidth}
    \input{./plots/angle_20dB.pgf}
    \caption{\label{fig:angle_20dB} Average of $|\phi_k|^2$}
  \end{subfigure}
      \par\medskip
  \begin{subfigure}[b]{\textwidth}
    \centering\input{./plots/legend_3col.pgf}
  \end{subfigure}
  \caption{\label{fig:20dB} Results for the i.i.d.\ channel model
  %with $\|\mathbf{H}\|_2^2$
  at $\rho_e = \rho_o = 20$ dB with $M_r = 4, M_t = 32$}
\end{figure*}

From Figs.~\ref{fig:m10dB}-\ref{fig:20dB},
we make the following remarks:
\begin{enumerate}
\item
There is a minor performance gap (both in terms of gains and angles) between
the optimal and suboptimal variants of the SLS estimator across all the three
SNRs, even though there is a significant complexity reduction with the suboptimal
variant.
Thus, this study motivates the use of the suboptimal variant of the SLS estimator
over the optimal variant.

\item
In the low-SNR regime typical of mmWave settings, the summed power method
significantly outperforms all the methods for small $k$ values, whereas the
additional channel estimation step of the SLS estimator contributes to its
utility for large $k$ values. While the method from~\cite{gazor_communications_2010}
is better in performance than the one from~\cite{dahl_blind_2004}, neither method
produces a performance comparable to the schemes proposed in this work.

\item
The performance of the schemes in~\cite{gazor_communications_2010} and~\cite{dahl_blind_2004}
improve with SNR. In the high-SNR regime, both methods become comparable to the
SLS estimator. However, the summed power method is significantly inferior in this
regime as it cannot suppress the effect of noise from the beamformer
estimates for large $k$ values.

\item
The switching between the SLS part and the summed power part means that the
LISP method shows a switch in terms of
performance at $k = k_{\sf switch} = \max(M_r, M_t) = 32$. But more
importantly, in the low-SNR regime, the LISP method approaches the
performance of the summed power method for large $k$ and in the high-SNR regime,
it approaches the performance of the SLS estimator (even for small $k$ values)
without the additional complexity overhead of these methods. Thus, this
method may be a suitable low-complexity alternative to the SLS estimator in the
medium- to high-SNR regime.

\item
In the sparse mmWave setting with low SNR, the summed power method outperforms
all the methods over all the values of $k$ considered here. The LISP method
quickly approaches the performance of the summed power method after
$k = k_{\sf switch}$.
\end{enumerate}

Summarizing the above statements, we have the following conclusions: i) In the
low-SNR regime, the summed power method is advantageous for small $k$ and the
SLS estimator is advantageous for large $k$. If computational complexity is an
important issue for large $k$, the LISP method can be a useful alternative.
ii) In the high-SNR regime, the LISP method or the method proposed
in~\cite{dahl_blind_2004} are advantageous for all $k$. iii) These broad
conclusions appear to be true for both i.i.d.\ as well as sparse mmWave
channel models.

\subsection{Impact of Antenna Dimensions and $k_{\sf switch}$}

Fig.~\ref{fig:gain_vs_antennas} studies the impact of $M_t$ (as $M_t$ increases
from $6$ to $64$) on the effective channel gain after $k=100$ iterations with
the different beam alignment techniques. The low-SNR regime corresponding to
$\rho_e = \rho_o = -10$ dB and $M_r = 4$ is considered in this study.
Figs.~\ref{fig:gain_vs_antennas_iid} and~\ref{fig:gain_vs_antennas_sparse}
present the results for the i.i.d.\ channel model and the sparse mmWave channel
model introduced earlier.

This study reinforces the advantages of the summed power and least-squares initialized
summed power methods relative to other methods. In particular, the
performance of the summed power method remains approximately invariant in the
i.i.d.\ case as $M_t$ increases.  On the other hand, the smaller rank of the
channel matrix in the sparse case improves the fraction of power in the
dominant eigen-mode, which is reflected in improving performance as $M_t$
increases. But more importantly, the performance of all other schemes depreciate
with $M_t$ suggesting their sensitivity to larger antenna dimensions. Nevertheless,
the LISP method appears closest to the summed power method in performance at
low-SNR and is also superior at high-SNRs. From these results, we conclude that
the proposed beam alignment techniques and in particular, the LISP method can
deliver substantial performance improvement as $M_t$ increases with low
complexity and feedback overheads making them viable candidates for practical
large/massive MIMO systems.

Figs.~\ref{fig:0dB_kswitch} and~\ref{fig:m10dB_kswitch} study the choice of
$k_{\sf switch}$ to be used in the LISP method with $M_r = 4$, $M_t = 32$ and
$\rho_e = \rho_o = 0$ dB and $\rho_e = \rho_o = -10$ dB, respectively. From
Fig.~\ref{fig:0dB_kswitch}, we note that there exists an optimal $k_{\sf switch}$
that maximizes the effective channel gain for both the i.i.d.\ and sparse mmWave
channel models. The optimal $k_{\sf switch}$ value is typically small in the case of
sparse mmWave channels for both SNR settings. In fact, for $\rho_e = \rho_0 = -10$ dB,
the optimal $k_{\sf switch}$ in the sparse setting is $1$ implying that the summed
power method starting at $k = 1$ is better than a noisy initialization based on the
SLS estimator. While the optimal $k_{\sf switch}$ can be high in the i.i.d.\ setting,
constraining it to be a small number does not result in a significantly poorer
performance relative to the optimal $k_{\sf switch}$ value. Thus,
Figs.~\ref{fig:0dB_kswitch} and~\ref{fig:m10dB_kswitch} suggest that, in the
moderate- to high-SNR regime and depending on the level of richness/sparsity structure
of the channel, a small $k_{\sf switch}$ may be a better choice than the use of summed
power method ($k_{\sf switch} = 1$). Thus, an improved performance can be ensured with the
LISP method at the cost of a small feedback and complexity overhead.

\subsection{Comparison with a Pilot-Based Channel Estimation Scheme}
\label{sec:comp-with-pilot}
We are now interested in comparing the performance of the proposed beam alignment schemes
with a traditional pilot-based channel estimation scheme. In order to simplify the structure
of the pilot-based scheme, we assume that the channel matrices are i.i.d.\ Rayleigh fading.
In order to fairly compare the iterative schemes with the batch-oriented pilot-based scheme,
we impose a constraint on the total energy used during the beam alignment/channel estimation
phase. Let $k_{\mathsf{max}}$ be the number of time slots allocated for this phase. With the
iterative schemes considered in this work, the total energy used by nodes 1 and 2 reduces to
$\rho_{o}\cdot k_{\mathsf{max}}$ and $\rho_{e}\cdot k_{\mathsf{max}}$, respectively. With the
pilot-based scheme, it is well understood~\cite{medard,hassibi,raghavan_sparse1} that the
quality of the channel estimate {\em only} depends on the energy in the training matrices
(denoted as $\mathbf{P}_{o}$ and $\mathbf{P}_{e}$ for the downlink and uplink, respectively)
as long as the number of pilot symbols exceeds the transmit antenna dimensions. Thus, we can
assume that $\mathbf{P}_{o}$ and $\mathbf{P}_{e}$ are $M_t \times M_t$ and $M_r \times M_r$
scaled-unitary matrices meeting the energy constraint, respectively. With the energy scaling,
we have the following system equations:
\begin{IEEEeqnarray}{rCl}
  \label{eq:mmse_downlink_io}
  \mathbf{Y}_{o} = \sqrt{\rho_{o} \cdot k_{\mathsf{max}}/M_{t}}\, \mathbf{H}\mathbf{P}_{o}
  + \mathbf{N}_{o}
\end{IEEEeqnarray}
for the downlink, and
\begin{IEEEeqnarray}{rCl}
  \label{eq:mmse_uplink_io}
  \mathbf{Y}_{e} = \sqrt{\rho_{e} \cdot k_{\mathsf{max}}/M_{r}}\, \mathbf{H}^{\T}\mathbf{P}_{e} + \mathbf{N}_{e}
\end{IEEEeqnarray}
for the uplink.

Upon reception of $\mathbf{Y}_{o}$ and $\mathbf{Y}_{e}$, each node computes a minimum mean-squared
error (MMSE) channel estimate as follows:
\begin{IEEEeqnarray}{rCl}
\widehat{\mathbf{H}}_o & = & \frac{  \sqrt{\rho_{o} \cdot k_{\mathsf{max}}/M_{t}} }
{ 1 + \rho_{o} \cdot k_{\mathsf{max}}/M_{t} } \cdot \mathbf{Y}_{o} \mathbf{P}_{o}^*
\\
\widehat{\mathbf{H}}_e & = & \frac{  \sqrt{\rho_{e} \cdot k_{\mathsf{max}}/M_{r}} }
{ 1 + \rho_{e} \cdot k_{\mathsf{max}}/M_{r} } \cdot \mathbf{Y}_{e} \mathbf{P}_{e}^*.
\end{IEEEeqnarray}
The beamformers are estimated using the SVD of the channel estimates. In our study, we
use scaled discrete Fourier transform (DFT) matrices for $\mathbf{P}_{o}$ and $\mathbf{P}_{e}$
over the i.i.d.\ channel. With $k_{\mathsf{max}} = 100$, the normalized channel gain
across different SNR values is plotted for the different schemes in Fig.~\ref{fig:gain_vs_snr_iid}.
These results show that in addition to outperforming iterative schemes from prior works in the
low-SNR regime, the proposed methods also compare favorably to the pilot-based channel estimation
scheme. The pilot-based scheme requires a substantial pre-beamforming SNR (over $5$-$10$ dB)
for improved performance which may not be feasible in practical mmWave systems. Further, it
also requires a computational overhead in computing the SVD of the channel estimate.

\section{Concluding Remarks}
\label{sec:conclusion}
This paper studied the problem of estimating the dominant singular vectors of a
MIMO channel matrix in a TDD system. Such a task is of importance in realizing
the full analog beamforming gains in practical mmWave systems, typically
impaired with low SNR. We presented multiple iterative approaches based on the
power method to address this problem. These approaches included batch and
sequential least-squares estimation, summed power method, and least-squares
initialized summed power method. Numerical studies and analysis
established that the proposed approaches enjoy several advantages over
competing approaches from the literature. These advantages include improved
convergence and/or performance (beamforming gain) at low- as well as high-SNR
at a low-complexity and feedback overhead.

That said, this paper has only scratched the surface of the noisy beam alignment
problem. Further studies on developing an analytical/manifold optimization-based
framework for the rate of convergence of the proposed algorithms as a function of
the SNR, antenna dimensions, mmWave channel eigen-mode/sparsity structure, etc.\
are important. Such a step could also be of independent interest in problems in
machine learning, principal component analysis, and linear algebra. Other problems
of interest include understanding the impact of an imperfect (e.g., finite-rate,
noisy, etc.) feedback link on the performance of the sequential least-squares
estimation scheme, performance comparison with other directional learning
approaches~\cite{raghavan_jstsp}, impact of temporal variation in the channel
and wideband aspects on the performance of the proposed schemes, extending the
proposed analog beamforming schemes to a hybrid architectural set-up or
multi-user settings~\cite{vasanth_gcom16}, intuitive understanding of $k_{\sf switch}$
and further optimization of the beam alignment parameters given an asymmetrical
antenna setting in the single-user case, etc.

\appendix
\subsection{Proof of Theorem~\ref{thm_1}}
\label{appA}
%\begin{IEEEproof}
 The derivation of~\eqref{eq:Ho_batch_cov} mirrors~\eqref{eq:He_batch_cov}
 and thus it suffices to establish~\eqref{eq:He_batch_cov}.
  Transposing~\eqref{eq:slot2_matrix} at time $k-1$, we get
  \begin{IEEEeqnarray}{rCl}
    \label{eq:lemma_cov_transpose}
    \bY_{e, k-1}^{\T} = \sqrt{\rho_{e}}\,\bZ_{k-1}^{*}\bH + \bN_{e, k-1}^{\T}.
  \end{IEEEeqnarray}
  Since the columns of $\bH$ are i.i.d.\
  complex Gaussian random vectors, we focus on the first column without loss in
  generality. Let this first column of $\bH$ be denoted as $\bh_{1}$ and let its
  estimator be $\widehat{\bh}_{1}$. With %$\bm{\gamma}$
  $\widetilde{ {\mathbf y}}$ denoting the first column
  of $\bY_{e, k-1}^{\T}$, we have
  \begin{IEEEeqnarray}{rCl}
    \label{eq:lemma_cov_firstcol}
    %\bm{\gamma}
    \widetilde{ {\mathbf y}} = \sqrt{\rho_{e}}\,\bZ_{k-1}^{*}\bh_{1} + \widetilde{\bn},
  \end{IEEEeqnarray}
  where $\widetilde{\bn}$ is the first column of $\bN_{e,k-1}^{\T}$ with
  i.i.d.\ complex Gaussian entries. The estimator of $\bh_{1}$ is given as
  \begin{IEEEeqnarray}{rCl}
    \label{eq:lemma_cov_firstcol_estimator}
    \widehat{\bh}_{1} = \frac{ \left(\bZ_{k-1}^{*} \right)^ {\dagger}\widetilde{ {\mathbf y}} }
    %\bm{\gamma}}
    {\sqrt{\rho_{e}}}
  \end{IEEEeqnarray}
  with error covariance matrix $\bC_{e, k}$, defined as,
    $\bC_{e, k} \triangleq \Exp{\left(\bh_{1} - \widehat{\bh}_{1} \right)\left( \bh_{1} - \widehat{\bh}_{1}
    \right)^{*}}.$

  It can be seen that
  \begin{IEEEeqnarray}{rCl}
    \label{eq:lemmma_cov_halfterm}
      \bh_{1} - \widehat{\bh}_{1} &=&
      \bh_{1} -
      \frac{ \left( \bZ_{k-1}^{*} \right)^{\dagger}}{\sqrt{\rho_{e}}}\left( \sqrt{\rho}_{e}\bZ_{k-1}^{*}\bh_{1} +\widetilde{\bn} \right) %\IEEEnonumber
    \\
    &=& \frac{- \left( \bZ_{k-1}^{*} \right)^{\dagger}}{\sqrt{\rho_{e}}} \widetilde{\bn}
  \end{IEEEeqnarray}
  and
  \begin{IEEEeqnarray}{rCl}
    \label{eq:lemma_cov_plugin}
    \bC_{e, k} &=& \frac{ \left( \bZ_{k-1}^{*} \right)^{\dagger}}{\sqrt{\rho_{e}}} \Exp{\widetilde{\bn}\widetilde{\bn}^{*}} \frac{ \left( \bZ_{k-1} \right)^{\dagger}}{\sqrt{\rho_{e}}} %\IEEEnonumber
    \\
    &=& \frac{1}{\rho_{e}} \left( \mathbf{Z}_{k-1} \mathbf{Z}_{k-1}^* \right)^{-1} \mathbf{Z}_{k-1} \mathbf{Z}_{k-1}^* \left( \mathbf{Z}_{k-1} \mathbf{Z}_{k-1}^* \right)^{-1} %\IEEEnonumber
    \\
    &=& \frac{1}{\rho_{e}} \left( \mathbf{Z}_{k-1} \mathbf{Z}_{k-1}^* \right)^{-1}. \hspace*{\fill} %\IEEEQEDhere\
  \end{IEEEeqnarray}
%\end{IEEEproof}
Note that the above equation holds only under the i.i.d.\ ${\bf H}$ assumption and for $k \geq M_r$.
\qed

\begin{figure*}
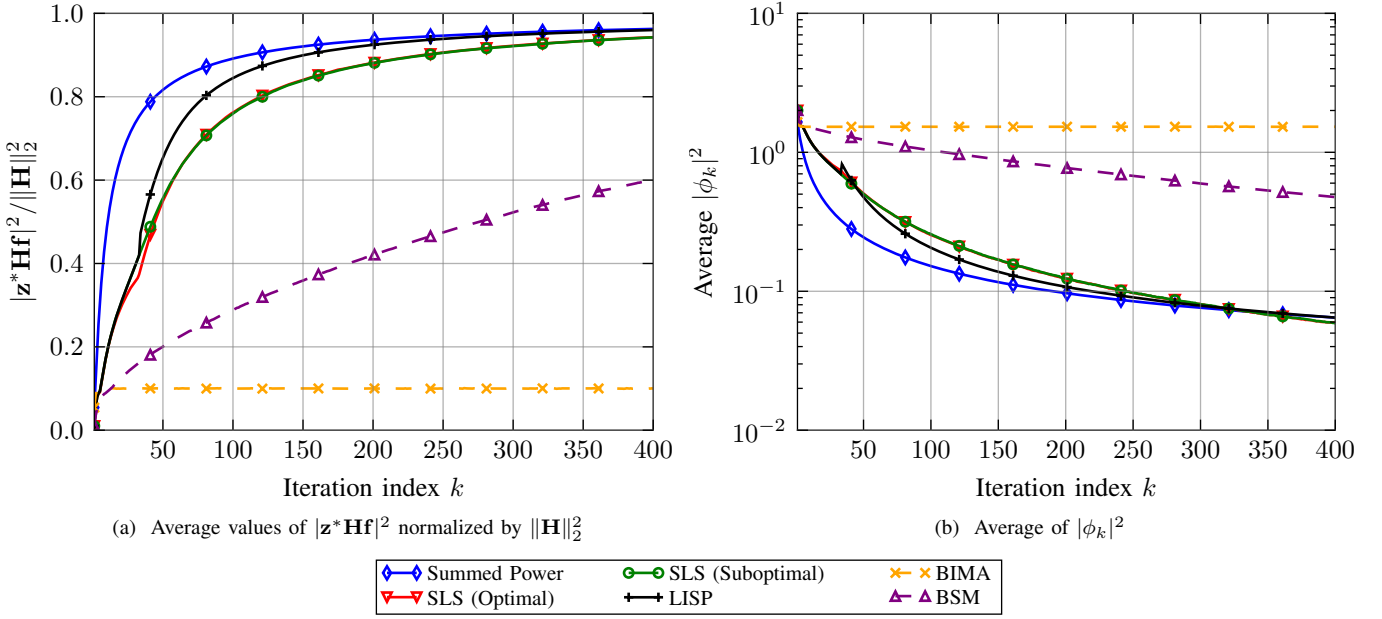

  \centering
  \begin{subfigure}[htb!]{.5\textwidth}
    \input{./plots/gain_m10dB_sparse.pgf}
    \caption{\label{fig:gains_m10dB_sparse} Average values of
      $|\mathbf{z}^* \mathbf{H} \mathbf{f}|^2$ normalized by
      $\|\mathbf{H}\|_2^2$}
  \end{subfigure}\hfill
  \begin{subfigure}[htb!]{.5\textwidth}
    \input{./plots/angle_m10dB_sparse.pgf}
    \caption{\label{fig:angle_m10dB_sparse} Average of $|\phi_k|^2$}
  \end{subfigure}
      \par\medskip
  \begin{subfigure}[htb!]{\textwidth}
    \centering\input{./plots/legend_3col.pgf}
  \end{subfigure}
  \caption{\label{fig:m10dB_sparse} Results for the sparse mmWave channel model at
  $\rho_e = \rho_o = -10$ dB with $M_r = 4, M_t = 32$}
\end{figure*}

\begin{figure*}
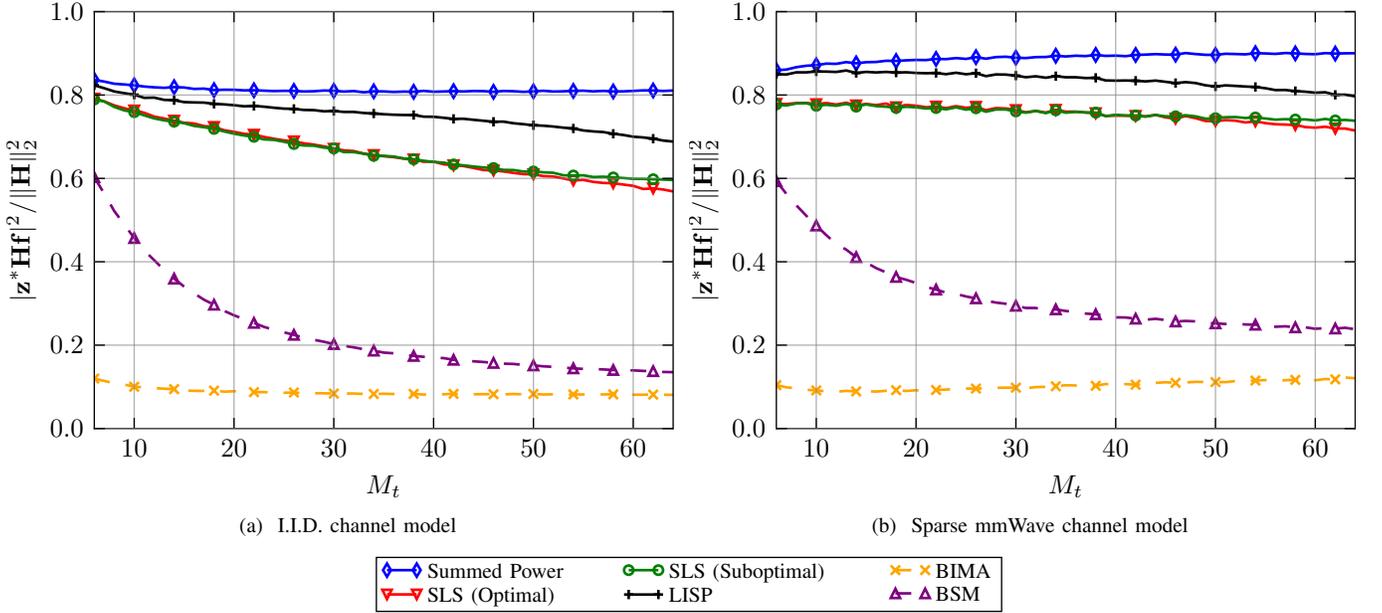

  \centering
  \begin{subfigure}[t]{.5\textwidth}
    \input{./plots/gain_vs_antennas_iid_m10dB_10000runs_mean.pgf}
    \caption{\label{fig:gain_vs_antennas_iid} I.I.D.\ channel model}
  \end{subfigure}\hfill
  \begin{subfigure}[t]{.5\textwidth}
    \input{./plots/gain_vs_antennas_sparse_m10dB_10000runs_mean.pgf}
    \caption{\label{fig:gain_vs_antennas_sparse} Sparse mmWave channel model}
  \end{subfigure}
      \par\medskip
  \begin{subfigure}[b]{\textwidth}
    \centering\input{./plots/legend_3col.pgf}
  \end{subfigure}
  \caption{\label{fig:gain_vs_antennas}Average value of
    $|\mathbf{z}^* \mathbf{H} \mathbf{f}|^2$ normalized by $\|\mathbf{H}\|_2^2$
    at $k=100$ for $\rho_e = \rho_o = -10$ dB using
    different channel models with $M_{r} = 4$ and $M_{t} \in {6, 8, 10, \cdots, 64}$.}
\end{figure*}

\begin{figure}[h!]
  \centering
  \input{./plots/HybridKswitch_100_its_84000_runs_0dB_4x32.pgf}
  \caption{\label{fig:0dB_kswitch} Normalized channel gain of the
  LISP method as a function of $k_{\mathsf{switch}}$ with $M_r = 4,
  M_{t} = 32$, $\rho_{e} = \rho_{o} = 0$ dB and $k_{\mathsf{max}} = 100$.
    }
\end{figure}

\begin{figure}[h!]
  \centering
  \input{./plots/HybridKswitch_400_its_84000_runs_m10dB_4x32.pgf}
  \caption{\label{fig:m10dB_kswitch} Normalized channel gain of the LISP method
    as a function of $k_{\mathsf{switch}}$ with $M_r = 4, M_{t} = 32$,
    $\rho_{e} = \rho_{o} = -10$ dB and $k_{\mathsf{max}} = 400$.
    }
\end{figure}

\begin{figure}[ht]
  \centering
  \input{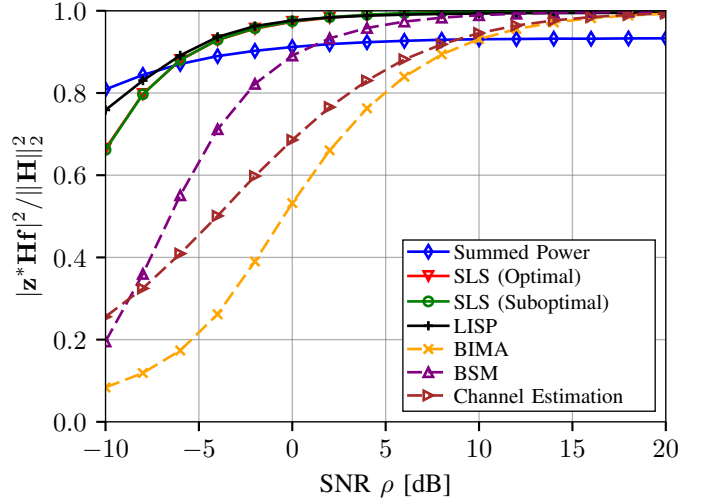}
  \caption{\label{fig:gain_vs_snr_iid} Normalized channel gain with
  $k_{\mathsf{max}}=100$ for varying values of $\rho = \rho_o = \rho_e$
  in the i.i.d.\ Rayleigh fading channel case.}
\end{figure}

\subsection{Proof of Theorem~\ref{lem:seq_equivalent}}
\label{appC}
  Without loss in generality, we can assume that $\rho_{e} = \rho_{o} = 1$.
  From~\eqref{eq:Ho_batch}, we have
  \begin{align}
  %\begin{IEEEeqnarray*}{rCl}
    & \widehat{\mathbf{H}}_{o,k}^{\mathrm{Batch}}
    \nonumber \\
    &= \mathbf{Y}_{o,k} \mathbf{F}_{k}^{*} \mathbf{C}_{o, k} = \mathbf{Y}_{o,k} \mathbf{F}_{k}^{*}\left( \mathbf{F}_{k} \mathbf{F}_{k}^{*} \right)^{-1} \\
    &= \begin{bmatrix} \mathbf{Y}_{o, k-1} \ \ \mathbf{y}_{o}[k] \end{bmatrix} \begin{bmatrix} \mathbf{F}_{k-1}^{*} \\ \mathbf{f}^{*}[k] \end{bmatrix} \left( \begin{bmatrix} \mathbf{F}_{k-1} & \mathbf{f}[k] \end{bmatrix} \begin{bmatrix} \mathbf{F}_{k-1}^{*} \\ \mathbf{f}^{*}[k] \end{bmatrix} \right)^{-1} \\
    &= \begin{bmatrix}\mathbf{Y}_{o, k-1} \mathbf{F}_{k-1}^{*} + \mathbf{y}_{o}[k] \mathbf{f}^{*}[k] \end{bmatrix} \left( \mathbf{F}_{k-1} \mathbf{F}_{k-1}^{*} + \mathbf{f}[k] \mathbf{f}^{*}[k] \right)^{-1}.
   %\end{IEEEeqnarray*}
   \end{align}

  Substituting~\eqref{eq:Ho_batch_cov} and applying the Woodbury matrix
  identity~\cite{horn_matrix_2012} to the second term, we get
  \begin{IEEEeqnarray}{rCl}
    \mathbf{C}_{o, k} = \mathbf{C}_{o, k-1} - \frac{\mathbf{C}_{o, k-1} \mathbf{f}[k] \mathbf{f}^{*}[k] \mathbf{C}_{o, k-1} }{1 +  \mathbf{f}^{*}[k]\mathbf{C}_{o, k-1} \mathbf{f}[k]}.
  \end{IEEEeqnarray}
  We now let
  \begin{align}
  %\begin{IEEEeqnarray}{rCl}
    \mathbf{K}_{o, k} = \frac{\mathbf{f}^{*}[k] \mathbf{C}_{o, k-1} }{1 +  \mathbf{f}^{*}[k]\mathbf{C}_{o, k-1} \mathbf{f}[k]}
  %\end{IEEEeqnarray}
  \end{align}
  and write
  \begin{align}
  %\begin{IEEEeqnarray*}{rCl}
    \widehat{\mathbf{H}}_{o, k}^{\mathrm{Batch}} &= \mathbf{Y}_{o, k-1} \mathbf{F}_{k-1}^{*}
    \mathbf{C}_{o, k-1} - \mathbf{Y}_{o, k-1} \mathbf{F}_{k-1}^{*} \mathbf{C}_{o, k-1} \mathbf{f}[k]
    \mathbf{K}_{o,k} \nonumber \\
    & + \mathbf{y}_{o}[k] \mathbf{f}^{*}[k] \mathbf{C}_{o, k-1} - \mathbf{y}_{o}[k] \mathbf{f}^{*}[k] \mathbf{C}_{o, k-1} \mathbf{f}[k] \mathbf{K}_{o, k}.
  %\end{IEEEeqnarray*}
  \end{align}
  Now, since
  \begin{align}
  %\begin{IEEEeqnarray}{rCl}
    \mathbf{y}_{o}[k] \mathbf{f}^{*}[k] \mathbf{C}_{o, k-1} = \mathbf{y}_{o}[k] \left( 1 +  \mathbf{f}^{*}[k]\mathbf{C}_{o, k-1} \mathbf{f}[k] \right) \mathbf{K}_{o, k},
  %\end{IEEEeqnarray}
  \end{align}
  we get
  \begin{IEEEeqnarray}{rCl}
    \widehat{\mathbf{H}}_{o, k}^{\mathrm{Batch}} &=& \widehat{\mathbf{H}}_{o, k-1} \left( \mathbf{I} - \mathbf{f}[k] \mathbf{K}_{o, k} \right) + \mathbf{y}_{o}[k] \mathbf{K}_{o, k} \\
    &=& \widehat{\mathbf{H}}_{o, k-1} + \left( \mathbf{y}_{o}[k] - \widehat{\mathbf{H}}_{o, k-1} \mathbf{f}[k] \right) \mathbf{K}_{o, k} \\
    &=& \widehat{\mathbf{H}}_{o, k}^{\mathrm{Seq}}. %\hspace*{\fill}\IEEEQEDhere\
  \end{IEEEeqnarray}
  \qed

\subsection{Proof of Theorem~\ref{lem:seq_non_optimal}}
\label{app:lem:seq_non_optimal}

Along the same lines of the proof of Theorem~\ref{lem:seq_equivalent}, let us
assume that $\rho_{e}~=~\rho_{o}~=~1$. Suppose that node 2 has access to
$M_{t}$ previous observations at time slot $k=0$, indexed from $k=-(M_{t}-1)$ to
$k=0$. Using this data, node 2 could thus compute the batch estimate at time
slot $k=0$, given as
\begin{IEEEeqnarray}{rCl}
  \widehat{\mathbf{H}}_{o,0}^{\mathrm{Batch}} &&= \mathbf{Y}_{o,0} \mathbf{F}_{0}^{\dagger}
  %\\ &&
  =\mathbf{Y}_{o,0} \mathbf{F}_{0}^*\left( \mathbf{F}_{0} \mathbf{F}_{0}^* \right)^{-1},
\end{IEEEeqnarray}
where
$\mathbf{F}_{0} = \begin{bmatrix}\mathbf{f}[-(M_{t}-1)] & \mathbf{f}[-M_{t}] &
  \hdots & \mathbf{f}[0] \end{bmatrix}$ and
$\mathbf{Y}_{o,0} = \begin{bmatrix}\mathbf{y}_{o}[-(M_{t}-1)] &
  \mathbf{y}_{o}[-M_{t}] & \hdots & \mathbf{y}_{o}[0] \end{bmatrix}$. Using
Theorem~\ref{thm_1}, we note that the covariance matrix of each column of
this estimated matrix is given as
\begin{IEEEeqnarray}{c}
  \mathbf{C}_{o,0} = \left( \mathbf{F}_{0} \mathbf{F}_{0}^* \right)^{-1}.
\end{IEEEeqnarray}
Applying the result of Theorem~\ref{lem:seq_equivalent}, we note that for any
$k > 0$, a sequential least-squares estimator would be identical to the batch
estimator using all of the data from $k=-(M_{t}-1)$ up to $k$. We can thus write
%\begin{IEEEeqnarray}{rCl}
\begin{align}
  \label{eq:prop:seq_init:seq_batch}
  \widehat{\mathbf{H}}_{o,k}^{\mathrm{Seq}} & =  \widehat{\mathbf{H}}_{o,k}^{\mathrm{Batch}}
  = \mathbf{Y}_{o,k} \mathbf{F}_{k}^{\dagger}
  \\
  & =  \left( \sum_{n=-(M_{t}-1)}^{k} \mathbf{y}_{o}[n] \mathbf{f}^*[n] \right)
  \left( \sum_{n=-(M_{t}-1)}^{k} \mathbf{f}[n] \mathbf{f}^*[n] \right)^{-1},
%\end{IEEEeqnarray}
\end{align}
where we have rewritten $\widehat{\mathbf{H}}_{o,k}^{\mathrm{Batch}}$ in terms
of individual vector outer products. After separating the hypothetical data
from $k=-(M_{t}-1)$ to $k=0$ from the data starting at $k=1$, we have for the
sequential estimator
\begin{align}
  \label{eq:prop:end}
  \widehat{\mathbf{H}}_{o,k}^{\mathrm{Seq}}
  & = \left( \sum_{n=-(M_{t}-1)}^{0} \mathbf{y}_{o}[n] \mathbf{f}^*[n] + \sum_{n=1}^{k} \mathbf{y}_{o}[n] \mathbf{f}^*[n] \right) \IEEEnonumber\\ &
  \cdot \left( \sum_{n=-(M_{t}-1)}^{0} \mathbf{f}[n] \mathbf{f}^*[n] + \sum_{n=1}^{k} \mathbf{f}[n] \mathbf{f}^*[n] \right)^{-1} \\
  & = \left( \mathbf{Y}_{o,0} \mathbf{F}_{0}^* + \mathbf{Y}_{o,k} \mathbf{F}_{k}^* \right) \left( \mathbf{F}_{0} \mathbf{F}_{0}^* + \mathbf{F}_{k} \mathbf{F}_{k}^* \right)^{-1} %\IEEEnonumber
  \\
  & = \left( \widehat{\mathbf{H}}_{o,0} \mathbf{C}_{o,0}^{-1} + \mathbf{Y}_{o,k} \mathbf{F}_{k}^* \right) \left( \mathbf{C}_{o, 0}^{-1} + \mathbf{F}_{k} \mathbf{F}_{k}^* \right)^{-1}.
\end{align}
Upon further inspection of~(\ref{eq:prop:end}), we observe that for any
$k>M_{t}$, the product $\mathbf{F}_{k} \mathbf{F}_{k}^*$ is invertible
and we can let $\mathbf{C}_{o, 0}^{-1}$ arbitrarily approach the matrix of all
zeros. This can be accomplished by choosing
$\mathbf{C}_{o, 0} = \alpha\mathbf{I}$. If $\mathbf{C}_{o, 0}$ is indeed chosen
this way,~(\ref{eq:prop:end}) loses its dependence on the previous data from the
supposition and we can start the sequential iteration at $k=1$. For
sufficiently large $\alpha$, the sequential estimator will approach the batch
estimator for any $k>M_{t}$.
\qed

\subsection{Proof of Lemma~\ref{lem:sum:diagonalization}}
\newcommand{\betak}{\beta_{k-1}}
\newcommand{\alphak}{\alpha_{k-1}}

\label{app:lem:sum:diagonalization}
Under Hypothesis 3, the state transition matrix for a real, diagonal
channel matrix
${\mathbf{H}}$ is given as
\begin{IEEEeqnarray}{rCl}
  \label{eq:app:derivation:stmatrix}
  \mathbf{S}_{k-1} = \begin{bmatrix}
    \mathbf{I} & \sqrt{\rho} \hsppp \betak \cdot {\mathbf{H}} \\ %\widetilde{\mathbf{H}}\\
    % \widetilde{\mathbf{H}}\alphak
   \sqrt{\rho} \hsppp \alphak \cdot {\mathbf{H}} & \mathbf{I}
  \end{bmatrix}
\end{IEEEeqnarray}
since ${\mathbf{H}} = {\mathbf{H}}^* = {\rm diag}\left([h_1, \cdots, h_M] \right)$.
Note that the size of ${\mathbf S}_{k-1}$ is $2M\times 2M$.
The characteristic equation of $\mathbf{S}_{k-1}$ is given as
%\begin{IEEEeqnarray}{rCl}
\begin{align}
  \label{eq:app:derivation:chareq}
  \chi(\mathbf{S}_{k-1}, \lambda) & =
  \mathrm{det}\left( \mathbf{S}_{k-1} - \lambda \mathbf{I} \right)
  %\nonumber
  \\
 & =
  \mathrm{det} \left( \begin{bmatrix}
      \mathbf{I} - \lambda \mathbf{I} & \sqrt{\rho} \hsppp \betak \cdot {\mathbf{H}} \\
      \sqrt{\rho} \hsppp \alphak \cdot {\mathbf{H}} %\widetilde{\mathbf{H}}\alphak
      & \mathbf{I} - \lambda \mathbf{I}
    \end{bmatrix} \right).
%\end{IEEEeqnarray}
\end{align}
By using the Schur complement lemma~\cite{horn_matrix_2012}, this equation
can be written as
\begin{align}
%\begin{IEEEeqnarray}{l}
&  \chi(\mathbf{S}_{k-1}, \lambda) = %\\
  \mathrm{det}\Big( \mathbf{I} - \lambda \mathbf{I} \Big) \cdot
  \nonumber \\
  & \label{eq:app:derivation:chareq_shur}
  {\hspace{0.3in}}
    \mathrm{det}\left( \mathbf{I} - \lambda \mathbf{I} - \rho\hsppp \alphak\betak \cdot {\mathbf{H}}
     \left( \mathbf{I} - \lambda \mathbf{I} \right)^{-1} {\mathbf{H}}
     \right).
%\end{IEEEeqnarray}
\end{align}
Since all of the matrices involved are diagonal, we can write the determinants
as the product of the diagonal elements, resulting in
%\begin{IEEEeqnarray}{rCl}
\begin{align}
  \label{eq:app:derivation:chareq_nodets}
  \chi(\mathbf{S}_{k-1},\lambda)
  &=
  \left( 1-\lambda \right)^{2M}
  \prod_{i=1}^{2M} \left\{ \left( 1-\lambda \right) - \frac{ \rho \hsppp \alphak \betak \cdot {h}_{i}^{2} }
  {1-\lambda} \right\}\\
  &=
  \prod_{i=1}^{2M} \left\{ \left( 1-\lambda \right)^{2} - \rho \hsppp \alphak \betak \cdot {h}_{i}^{2}
  \right\},
%\end{IEEEeqnarray}
\end{align}
which has $2M$ roots (denoted as $\lambda_1, \cdots, \lambda_{2M}$) of
the form $ %\lambda_i =
1 \pm \sqrt{\rho \hsppp \alphak\betak} \cdot {h}_{i} $
for $i=1, \hdots, M$. We can thus write the eigenvalue matrix
$\mathbf{\Lambda}_{k-1}$ as
\begin{IEEEeqnarray}{rCl}
  \mathbf{\Lambda}_{k-1} & = &
  \mathrm{diag} \left( \begin{bmatrix} \lambda_1 \\ \vdots \\
  \lambda_{2M} \end{bmatrix} \right)
  \\
  \label{eq:app:derivation:evmatrix}
  & = &
  \mathrm{diag} \left( \begin{bmatrix}
    1 + \sqrt{\rho \hsppp \alphak\betak} \cdot {h}_{1}  \\
    \vdots \\
     1 + \sqrt{\rho\hsppp \alphak\betak} \cdot {h}_{M}  \\
     1 - \sqrt{\rho \hsppp \alphak\betak} \cdot {h}_{1}  \\
     \vdots  \\
     1 - \sqrt{\rho\hsppp \alphak\betak} \cdot {h}_{M}
  \end{bmatrix} \right).
\end{IEEEeqnarray}
Solving for the $2M$ eigenvectors (i.e. solving
$\mathbf{S}_{k-1} {\mathbf{u}}_i  = \lambda_{i} \mathbf{u}_{i}$ for $i=1, \hdots, 2M$)
and normalizing each column to unit-norm finally results in the following eigenvector matrix:
\begin{IEEEeqnarray}{rCl}
  \label{eq:app:derivation:U}
  \mathbf{U}_{k-1} = \begin{bmatrix}
    \sqrt{\frac{\betak}{\alphak+ \betak}}\mathbf{I} & \sqrt{\frac{\betak}{\alphak+\betak}}\mathbf{I} \\
    \sqrt{\frac{\alphak}{\alphak+\betak}}\mathbf{I} & -\sqrt{\frac{\alphak}{\alphak+\betak}}\mathbf{I}
  \end{bmatrix}.
\end{IEEEeqnarray}
Note that ${\mathbf U}_{k-1}$ is not unitary in general and ${\mathbf S}_{k-1}$ can be
written as ${\mathbf S}_{k-1} = {\bf U}_{k-1} \cdot {\mathbf \Lambda}_{k-1} \cdot
\left({\mathbf U}_{k-1} \right)^{-1}$.
\qed

\subsection{Proof of Lemma~\ref{lem:alphak_betak}}
\label{app:lem:alphak_betak}

Let ${\bf H} = \{ {\bf H}_{m,n} \}$,
${\bf f}[i] = \left[f_{i,1}, \cdots, f_{i,M} \right]^{\T}$ and
${\bf z}[i] = \left[z_{i,1}, \cdots, z_{i,M} \right]^{\T}$. Also, let
${\bf n}_{e}[i] = \left[{\bf n}_{e,1}[i], \cdots, {\bf n}_{e,M}[i] \right]^{\T}$
and
${\bf n}_{o}[i] = \left[{\bf n}_{o,1}[i], \cdots, {\bf n}_{o,M}[i] \right]^{\T}$.
Now observe that
\begin{align}
\frac{1}{\alpha_k^2} & = \| {\mathbf s}_e[k] \|_2^2 %\\ & = &
=
\left\| \sum_{i=0}^k \conj{\mathbf{y}}_{e}[i] \right\|_2^2 \\
& = \sum_{n = 1}^M \left|
\sqrt{\rho} \sum_{m = 1}^M \conj{\bf H}_{m,n} \sum_{i = 0}^k z_{i,m}
+ \sum_{i = 0}^k \conj{\bf n}_{e,n}[i] \right|^2.
\label{eq:step6}
\end{align}
Similarly, we have
\begin{align}
\frac{1}{\beta_k^2} & = \sum_{m = 1}^M \left|
\sqrt{\rho} \sum_{n = 1}^M {\bf H}_{m,n} \sum_{i = 0}^k f_{i,n}
+ \sum_{i = 0}^k {\bf n}_{o,m}[i] \right|^2.
\label{eq:step7}
\end{align}

From~(\ref{eq:step6}) and~(\ref{eq:step7}), we have the following simplifications
\begin{eqnarray}
\frac{1}{ \rho \cdot \alpha_k^2} & \stackrel{(a)}{\approx} &
\sum_{n = 1}^M \left| \sum_{m = 1}^M \conj{\bf H}_{m,n} \sum_{i = 0}^k z_{i,m} \right|^2
\\
& \stackrel{(b)}{\approx} &
|{\sf C}_k|^2 \cdot \sum_{n = 1}^M \left| \sum_{m = 1}^M \conj{\bf H}_{m,n} \right|^2
\\
& \stackrel{(c)}{=} &
|{\sf C}_k|^2 \cdot {\sf Tr}({\bf H} {\bf H}^*)
\end{eqnarray}
where (a),~(b) and~(c) follow from Hypotheses 1-3, respectively. Similarly, we have
\begin{eqnarray}
\frac{1}{ \rho \cdot \beta_k^2} \approx |{\sf C}_k|^2 \cdot {\sf Tr}({\bf H} {\bf H}^*).
\end{eqnarray}
Thus, when Hypotheses 1-3 hold, $\alpha_k \approx \beta_k$ as $k$ increases.
\qed

\subsection{Proof of Theorem~\ref{lem:sum:sing_val_conv}}
\label{app:lem:sum:sing_val_conv}
When Hypotheses 1-3 hold, from Lemma~\ref{lem:alphak_betak}, we have that
$\alpha_k \approx \beta_k$ and ${\bf U}_{k-1} \approx \widetilde{\mathbf{U}}$.
Thus, the state-space model in~(\ref{eq:citebelow1}) can be written as
%It can then be shown that the state-space model using the diagonalized state
%transition matrix can be written as
%\begin{IEEEeqnarray*}{rCl}
\begin{align}
 & \mathbf{s}[k] =
 \nonumber \\
 & {\hspace{0.05in}}
 \widetilde{\mathbf{U}}
  \begin{bmatrix}
    \prod_{i=0}^{k-1} \left( \mathbf{I} +  \sqrt{\rho} \hsppp
    \alpha_i {\mathbf{H}} \right) & \mathbf{0} \\
    \mathbf{0} & \prod_{i=0}^{k-1} \left( \mathbf{I} - \sqrt{\rho} \hsppp
    \alpha_i {\mathbf{H}} \right)
  \end{bmatrix}
  \widetilde{\mathbf{U}}^{*} \mathbf{s}[0]
  \nonumber
  \\ %&&
  & {\hspace{-0.05in}} + \widetilde{\mathbf{U}} \sum_{\ell=1}^{k}
  \begin{bmatrix}
    \prod_{j=\ell}^{k-1} \left( \mathbf{I} +
    \sqrt{\rho} \hsppp \alpha_j {\mathbf{H}} \right) & \mathbf{0} \\
    \mathbf{0} &
    \prod_{j=\ell}^{k-1} \left( \mathbf{I} - \sqrt{\rho} \hsppp
    \alpha_j {\mathbf{H}} \right)
  \end{bmatrix}
  \widetilde{\mathbf{U}}^{*}\mathbf{n}[\ell]. %\IEEEnonumber
%\end{IEEEeqnarray*}
\label{eq:sum:sk_convergence}
\end{align}
Let us now consider the $2M\times 2M$-dimensional diagonal
matrix $\widetilde{\bf \Lambda}_{k-1,0}$
\begin{align}
%\begin{eqnarray}
%\begin{IEEEeqnarray*}{rCl}
  \label{eq:sum:Lambda}
  \widetilde{ \mathbf{\Lambda} }_{k-1,0} =
  \begin{bmatrix}
    \prod_{i=0}^{k-1} \left( \mathbf{I} +
    \sqrt{\rho} \hsppp \alpha_{i} {\mathbf{H}} \right) & \mathbf{0} \\
    \mathbf{0} & \prod_{i=0}^{k-1} \left( \mathbf{I} - \sqrt{\rho} \hsppp
    \alpha_{i} {\mathbf{H}} \right)
  \end{bmatrix}.
%\end{IEEEeqnarray*}
%\end{eqnarray}
\end{align}
From Hypothesis 3, since the diagonal entries of ${\bf H}$ are arranged in
non-increasing order and $h_1 > h_2$, we have
\begin{eqnarray}
\frac{ \prod_{i=0}^{k-1} (1 + \sqrt{\rho} \hsppp \alpha_i h_1) }
{ \prod_{i=0}^{k-1} (1 + \sqrt{\rho} \hsppp \alpha_i h_\ell) }
\approx \left( \frac{ h_1} { h_{\ell} } \right)^k \rightarrow \infty
{\hspace{0.05in}} {\sf as} {\hspace{0.05in}} k \rightarrow \infty
\end{eqnarray}
for $\ell = 2, \cdots , M$. Similarly, we have
\begin{eqnarray}
\frac{ \prod_{i=0}^{k-1} (1 - \sqrt{\rho} \hsppp \alpha_i h_1) }
{ \prod_{i=0}^{k-1} (1 - \sqrt{\rho} \hsppp \alpha_i h_\ell) }
\approx \left( \frac{ h_1} { h_{\ell} } \right)^k \rightarrow \infty
{\hspace{0.05in}} {\sf as} {\hspace{0.05in}} k \rightarrow \infty
\end{eqnarray}
for $\ell = 2, \cdots , M$ and
\begin{eqnarray}
\frac{ \prod_{i=0}^{k-1} (1 + \sqrt{\rho} \hsppp \alpha_i h_1) }
{ \prod_{i=0}^{k-1} (1 - \sqrt{\rho} \hsppp \alpha_i h_1) }
\approx (-1)^k {\hspace{0.05in}} {\sf as} {\hspace{0.05in}} k \rightarrow \infty.
\end{eqnarray}

Thus, the diagonal entries of $\widetilde{\bf \Lambda}_{k-1,0}$ are
dominated by (as $k$ increases) the first entry, which is denoted as
\begin{eqnarray}
\lambda_{1,0} \approx \left( \sqrt{\rho} \hsppp h_1 \right)^k \cdot
\prod_{i = 0}^k \alpha_i,
\end{eqnarray}
and the $(M+1)$th entry, which is denoted as
\begin{eqnarray}
\lambda_{M+1,0} \approx \left( - \sqrt{\rho} \hsppp h_1 \right)^k \cdot
\prod_{i = 0}^k \alpha_i.
\end{eqnarray}
Similarly, we can consider the diagonal
matrices $\widetilde{\bf \Lambda}_{k-1,\ell}$ for $\ell = 1, \cdots, k-1$:
\begin{align}
%\begin{eqnarray}
%\begin{IEEEeqnarray*}{rCl}
  %\label{eq:sum:Lambda}
  \widetilde{ \mathbf{\Lambda} }_{k-1,\ell} =
  \begin{bmatrix}
    \prod_{i=\ell}^{k-1} \left( \mathbf{I} +
    \sqrt{\rho} \hsppp \alpha_{i} {\mathbf{H}} \right) & \mathbf{0} \\
    \mathbf{0} & \prod_{i=\ell}^{k-1} \left( \mathbf{I} - \sqrt{\rho} \hsppp
    \alpha_{i} {\mathbf{H}} \right)
  \end{bmatrix}.
 % \end{eqnarray}
\end{align}
Following the same logic as before, these matrices are also dominated by
the first entry, which is denoted as
\begin{eqnarray}
\lambda_{1,\ell} \approx \left( \sqrt{\rho} \hsppp h_1 \right)^{k-\ell}
\cdot \prod_{i = \ell}^{k-1} \alpha_i,
\end{eqnarray}
and the $(M+1)$th entry, which is denoted as
\begin{eqnarray}
\lambda_{M+1,\ell} \approx \left( - \sqrt{\rho} \hsppp h_1 \right)^{k-\ell}
\cdot \prod_{i = \ell}^{k-1} \alpha_i.
\end{eqnarray}
With $\mathbf{s}[0] = \left[ s_1(0), \cdots, s_{2M}(0) \right]^{\T}$,
$\mathbf{s}[k] = \left[ s_1(k), \cdots, s_{2M}(k) \right]^{\T}$,
and
$\mathbf{n}[\ell] = \left[ n_1(\ell) , \cdots, n_{2M}(\ell) \right]^{\T}$,
it is straightforward to see that as $k$ increases and for $\rho \gg 1$, we
have
\begin{align}
%\begin{eqnarray}
\frac{ s_1(k) }{ \left( \sqrt{\rho} \hsppp h_1 \right)^{k}
\cdot \prod_{i = 0}^{k-1} \alpha_i }
& \rightarrow &
\left\{
\begin{array}{cc}
s_1(0) & {\sf if} {\hspace{0.05in}} k {\hspace{0.05in}} {\sf is}
{\hspace{0.05in}} {\sf even}
\\
s_{M+1}(0) & {\sf if} {\hspace{0.05in}} k {\hspace{0.05in}} {\sf is}
{\hspace{0.05in}} {\sf odd}
\end{array}
\right.
\\
\frac{ s_{M+1}(k) }{ \left( \sqrt{\rho} \hsppp h_1 \right)^{k}
\cdot \prod_{i = 0}^{k-1} \alpha_i }
& \rightarrow &
\left\{
\begin{array}{cc}
s_{M+1}(0) & {\sf if} {\hspace{0.05in}} k {\hspace{0.05in}} {\sf is}
{\hspace{0.05in}} {\sf even}
\\
s_{1}(0) & {\sf if} {\hspace{0.05in}} k {\hspace{0.05in}} {\sf is}
{\hspace{0.05in}} {\sf odd}.
\end{array}
\right.
%\end{eqnarray}
\end{align}
And we also have $\frac{ s_{\ell}(k) }
{ \left( \sqrt{\rho} \hsppp h_1 \right)^{k} \cdot
\prod_{i = 0}^{k-1} \alpha_i } \rightarrow 0$ for all $\ell \in
\{2 , \cdots, M, M+2 , \cdots, 2M \}$. Thus, ${\bf s}[k] \rightarrow
{\bf s}_{\sf opt}$ as $k$ increases.
\qed

\bibliographystyle{IEEEbib}
\bibliography{IEEEabrv,NBeamAlign_references}

% bios
\begin{IEEEbiography}%
  [{\includegraphics[width=1in,height=1.35in,clip,keepaspectratio]%
    {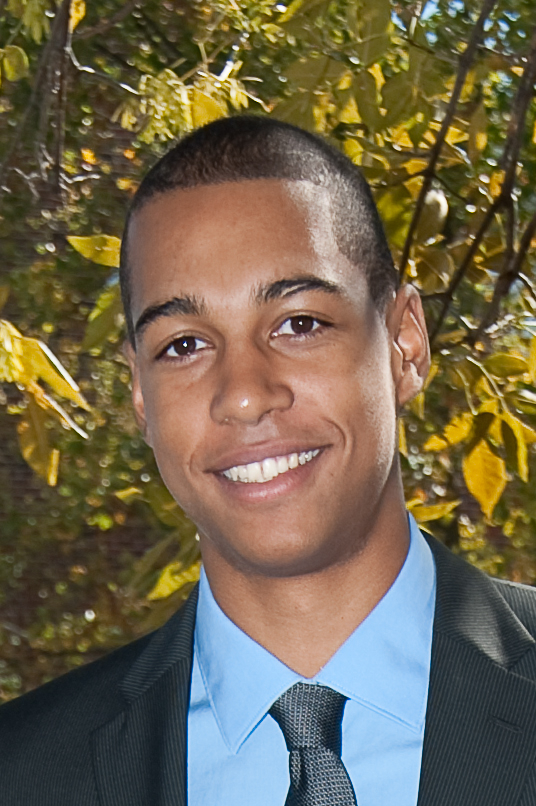}%
  }]{Dennis Ogbe (S'13)} received the B.S. degree (with honors) in electrical engineering in 2014 from Tennessee Technological University, Cookeville, TN and is currently working towards the Ph.D.\ degree at Purdue University, West Lafayette, IN. During the summer of 2016, he was an intern at Nokia Networks. His current research interests are in the design of adaptive multiple antenna wireless systems and software defined radio. Mr. Ogbe is an active member of Eta Kappa Nu.
\end{IEEEbiography}
\begin{IEEEbiography}%
  [{\includegraphics[width=1in,height=1.35in,clip,keepaspectratio]%
    {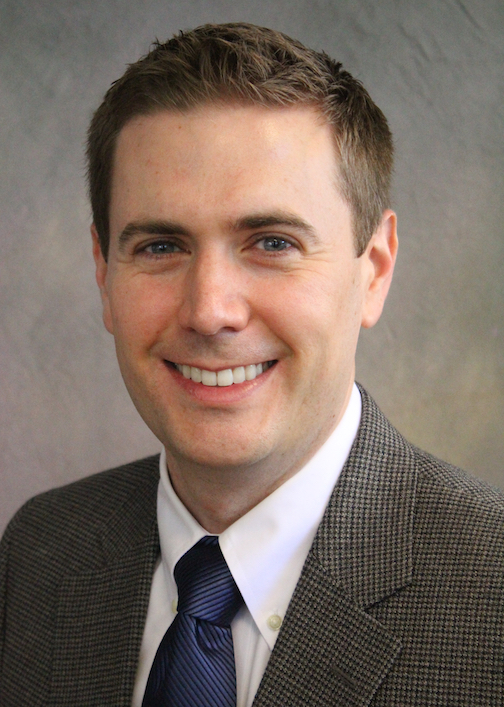}%
  }]{David J. Love (S'98--M'05--SM'09--F'15)} received the B.S. (with highest honors), M.S.E., and Ph.D. degrees in electrical engineering from the University of Texas at Austin in 2000, 2002, and 2004, respectively. Since August 2004, he has been with the School of Electrical and Computer Engineering, Purdue University, West Lafayette, IN, where he is now a Professor.  He has served as an Editor for the IEEE Transactions on Communications, an Associate Editor for the IEEE Transactions on Signal Processing, and a guest editor for special issues of the IEEE Journal on Selected Areas in Communications and the EURASIP Journal on Wireless Communications and Networking.  His industry experience includes work as a summer co-op and consultant for Texas Instruments.  Dr. Love holds 27 issued US patents.

Dr. Love was recognized as a Thomson Reuters Highly Cited Researcher in 2014 and 2015.   Along with his co-authors, he has won best paper awards from the IEEE Communications Society (2016 IEEE Communications Society Stephen O. Rice Prize), the IEEE Signal Processing Society (2015 IEEE Signal Processing Society Best Paper Award), and the IEEE Vehicular Technology Society (2009 IEEE Transactions on Vehicular Technology Jack Neubauer Memorial Award).  He has received multiple IEEE Global Communications Conference (Globecom) best paper awards.
\end{IEEEbiography}
\begin{IEEEbiography}%
  [{\includegraphics[width=1in,height=1.35in,clip,keepaspectratio]%
    {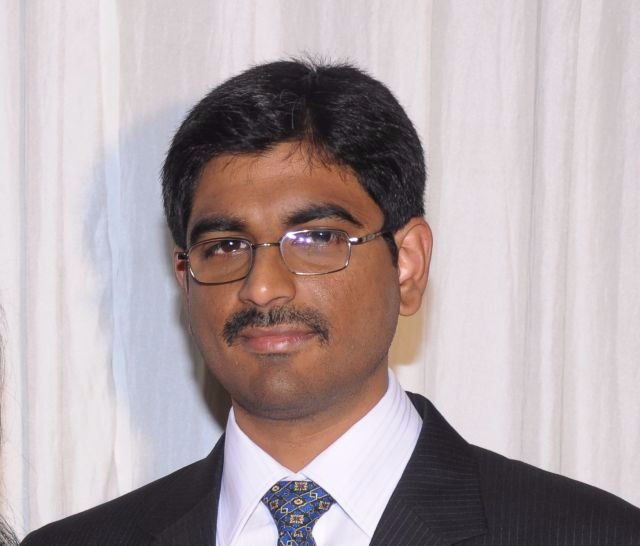}%
  }]{Vasanthan Raghavan (S'01--M'06--SM'11)} received the B.Tech degree in Electrical Engineering from the Indian Institute of Technology at Madras, India in 2001, M.S.\ and Ph.D.\ degrees in Electrical and Computer Engineering in 2004 and 2006, respectively, and M.A.\ degree in Mathematics in 2005, all from the University of Wisconsin, Madison, WI. He is currently with the New Jersey Research Center of Qualcomm, Inc. His research interests span multi-antenna communication techniques,information theory, quickest changepoint detection, and random matrix theory.
\end{IEEEbiography}

\end{document}